%% file: ms.tex
\documentclass[a4paper,10pt,fleqn]{article}

\usepackage{natbib}%
\usepackage[bottom=3cm,top=2cm,left=1.5cm,right=1.5cm]{geometry}%
\usepackage{framed,multirow,url,amsmath,amssymb,hyperref,setspace,graphicx,bm,xcolor,siunitx,nicefrac,multicol}
\usepackage[hang,centerlast]{subfigure}

\usepackage[english]{babel}
\usepackage[utf8]{inputenc}

\graphicspath{{./figures/},{./figures/png/}}%

\input{lbquad_definitions.tex}%

\definecolor{newcolor}{rgb}{.8,.349,.1}

\def\worktitle{Highly accurate numerical computation of implicitly defined volumes using the Laplace-Beltrami operator}%

\allowdisplaybreaks

\begin{document}
\title{\worktitle}%
\author{Johannes Kromer$^\dagger$ and Dieter Bothe$^\dagger$}%
\date{}%
\maketitle
\begin{center}
\small\em
\vspace{4mm}
$^\dagger$Mathematical Modeling and Analysis, Technische Universität Darmstadt\\ Alarich-Weiss-Strasse 10, 64287 Darmstadt, Germany\\Email for correspondence: \href{mailto:bothe@mma.tu-darmstadt.de}{bothe@mma.tu-darmstadt.de}%
\end{center}

\begin{abstract}
This paper introduces a novel method for the efficient and accurate computation of the volume of a domain whose boundary is given by an orientable hypersurface which is implicitly given as the iso-contour of a sufficiently smooth level-set function. After spatial discretization, local approximation of the hypersurface and application of the \name{Gaussian} divergence theorem, the volume integrals are transformed to surface integrals. Application of the surface divergence theorem allows for a further reduction to line integrals which are advantageous for numerical quadrature. We discuss the theoretical foundations and provide details of the numerical algorithm. Finally, we present numerical results for convex and non-convex hypersurfaces embedded in cuboidal domains, showing both high accuracy and thrid- to fourth-order convergence in space.
\end{abstract}
%

\section{Introduction}\label{sec:introduction}%
In the context of a two-phase flow problem in some bounded domain $\cell\subset\setR^d$ with $d\in\{2,3\}$, the spatial regions $\cell^\pm$ occupied by the respective phases, which are separated by an embedded hypersurface $\iface\subset\cell$, need to be easily identified. One way to achieve this consists in introducing a phase marker $f$ which, say, is 0 for $\vec{x}\in\cell^+$ and 1 for $\vec{x}\in\cell^-$, respectively. A spatial decomposition of the domain into pairwise disjoint cells $\cell_i$ allows to assign to each of those a fraction $f_i:=\lvert\cell_i\rvert^{-1}\int_{\cell_i}{f\dd{\vec{x}}}$ occupied by the first phase. While cells entirely confined in $\cell^\pm$ exhibit a marker value of one or zero, respectively, those intersected by the embedded hypersurface admit $0<f_i<1$. This representation provides the conceptual foundation of the well-known Volume-of-Fluid (VOF) method introduced by \citet{JCP_1981_vofm}. To solve an initial value two-phase flow problem, the above mentioned volume fractions $f_i$ need to be computed for a given domain and hypersurface. If accurate initial values are required, this task becomes particularly challenging for curved hypersurfaces, but also for seemingly simple\footnote{Simple in the sense that the description involves only a small set of parameters.} ones like ellipsoids. Thus, the objective of this work is to develop a numerical method for the accurate computation of those volume fractions. 

We first provide some relevant notation needed to precisely formulate the problem under consideration and to sketch the approach proposed in this work. The hypersurface $\iface\subset\cell$ induces a pairwise disjoint decomposition $\cell=\iface\cap\cell^+\cap\cell^-$, where we call $\cell^-$ and $\cell^+$ the \textit{interior} and \textit{exterior} subdomain, respectively. For the numerical approximation, the embedding domain $\cell$ is decomposed into a set of pairwise disjoint cells $\cell_i$, some of which are intersected by $\iface$, i.e.\ they contain patches $\iface_i:=\iface\cap\cell_i$ of the hypersurface. Note that $\iface=\bigcup\iface_i$. Any intersected cell again admits a disjoint decomposition into the hypersurface patch $\iface_i$, as well as an interior ($\cell_i^-$) and exterior ($\cell_i^+$) segment. The allocation property is inherited from the global decomposition of the embedding space, implying that, in a global sense, any $\vec{x}\in\cell_i\setminus\iface_i\subset\cell\setminus\iface$ is either interior or exterior. It is important to note that, locally, $\partial\iface_i\neq\emptyset$, even if the hypersurface is globally closed, i.e.\ $\partial\iface=\emptyset$. \refFig{illustration_notation} exemplifies the notation.


\begin{figure}[ht]
\null\hfill%
\includegraphics{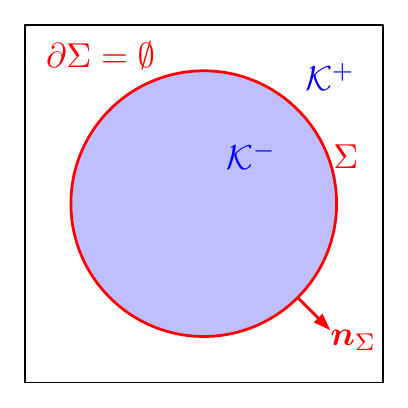}%
\hfill%
\includegraphics{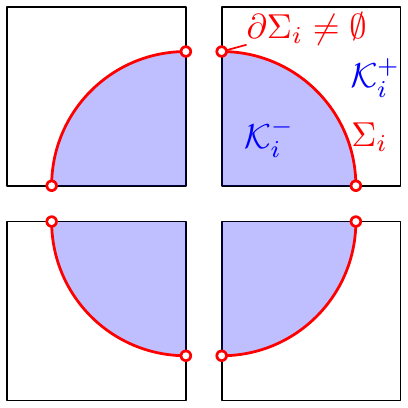}%
\hfill\null%
\caption{Illustration of the decomposition induced by a closed hypersurface $\iface$.}%
\label{fig:illustration_notation}%
\end{figure}

Henceforth we are concerned with a single intersected cell $\cell_i$ which is why we drop the cell index $i$ for ease of notation. The hypersurface patch $\iface\subset\cell\subset\setR^d$, with $d\in\{2,3\}$ denoting the spatial dimension, is assumed to be twice continuously differentiable with a simply connected, piecewise smooth boundary $\partial\iface\neq\emptyset$. Furthermore, the following assumptions are imposed: %
{\renewcommand{\theenumi}{\roman{enumi}}%
\begin{enumerate}
\item $\cell$ is convex with a boundary composed of planar polygons, $\partial\cell=\bigcup\mathcal{F}_k$. For technical simplicity, however, let $\cell=[0,1]^d$, implying that the cell faces $\mathcal{F}_k$ are rectangular. This assumption allows for a single parametrization of the boundary curve segment $\partial\iface_{k}=\iface\cap\mathcal{F}_k$. For general convex polyhedra the representation potentially requires a cumbersome piecewise definition. %
\item\label{num:vertexinterior} Both the interior and exterior segment contain at least one of the vertices of $\cell$, i.e.\ the hypersurface boundary $\partial\iface$ is not entirely contained in a single face $\mathcal{F}_k$. %
\item\label{num:connectivity} The division induced by the hypersurface yields simply connected sets $\cell^\pm$ and $\iface$, implying that $\cell$ contains a single patch of the hypersurface. This assumption resembles a resolution constraint to the underlying spatial discretization. %
\item For the principal curvatures $\curv_i$ it holds that $\curv_i d_\cell\lessapprox\num{e-2}$, where $d_\cell$ is a characteristic length of the cell $\cell$, e.g. the smallest edge length if $\cell$ is a cuboid. Note that this assumption actually is a resolution requirement. %
\end{enumerate}}
\refFig{cell_intersection_status} illustrates selected admissible and non-admissible setups.

\begin{figure}[h]
\null%
\hfill%
\subfigure[admissible setup]{\includegraphics[width=4cm]{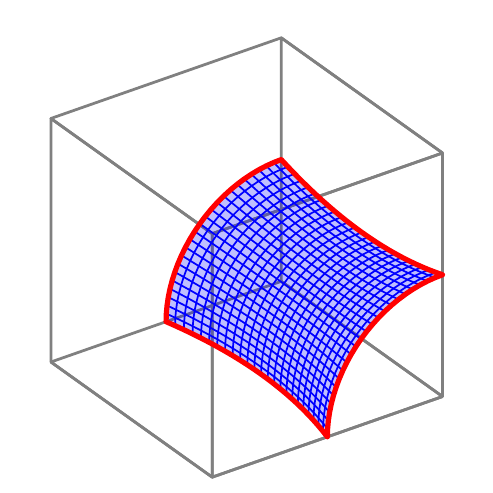}}%
\hfill%
\subfigure[\textbf{non}-admissible setup: the hypersurface patch $\iface$ is not connected (violation of \ref{num:connectivity})]{\includegraphics[width=4cm]{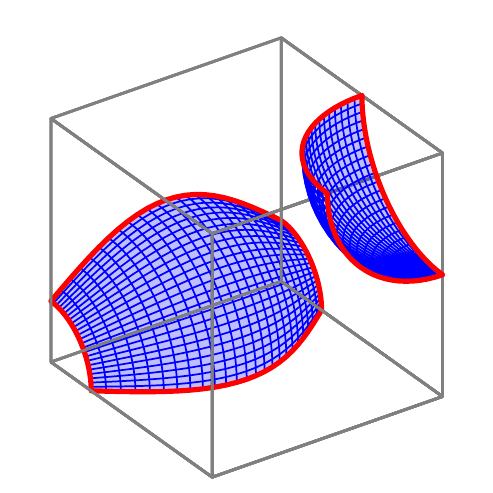}}%
\hfill%
\subfigure[\textbf{non}-admissible setup: all corners are located in the exterior/interior segment (violation of \ref{num:vertexinterior})]{\includegraphics[width=4cm]{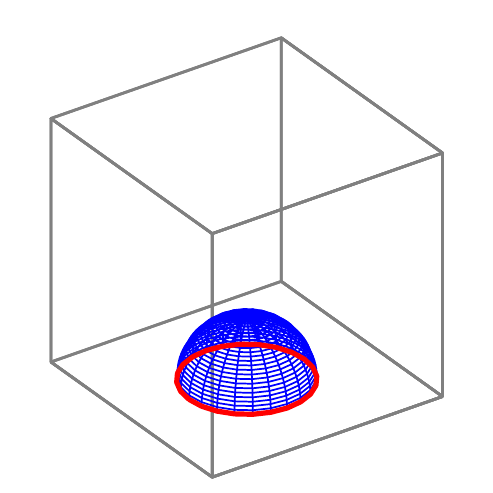}}%
\hfill%
\null%
\caption{Admissible and non-admissible intersection topologies of interface $\iface$ and cell $\cell$.}%
\label{fig:cell_intersection_status}%
\end{figure}
We are interested in the evaluation of
\begin{align}
\operatorname{vol}(\cell^-)&=\int\limits_{\cell^-}{1\,\dd{\vec{x}}}
\intertext{and employ the \name{Gaussian} divergence theorem to get}%
&=\frac{1}{3}\int\limits_{\partial\cell^-}{\iprod{\vec{x}}{\vec{n}}\,\dd{o}}=%
\frac{1}{3}\left[\int\limits_{\partial\cell^-\setminus\iface}{\iprod{\vec{x}}{\vec{n}_{\partial\cell^-}}\,\dd{o}}+\int\limits_{\iface}{\iprod{\vec{x}}{\ns}\,\dd{o}}\right],\label{eqn:volume_integration_base_equation}%
\end{align}
where $\iprod{\vec{a}}{\vec{b}}:=\vec{a}^{\mathsf{T}}\vec{b}$ is the standard inner product for $\vec{a},\vec{b}\in\setR^d$ and $\ns$ denotes the unit normal to $\iface$, pointing towards the exterior. Note that, by assumption, $\partial\cell^-\setminus\iface$ is a piecewise planar domain which considerably simplifies the numerical approximation of the associated integral. In contrast, the evaluation of the surface integral features some difficulties, one being the implicit definition of the integration domain itself. The key idea of the presented approach is the exploitation of the surface divergence theorem associated to $\iface$. For this purpose, assume for the moment that $u\in\hilbert[2]{\iface}$ is a given solution of \name{Laplace-Beltrami} equation
\begin{align}
\laplaces{u}=\iprod{\vec{x}}{\ns}\qquad\text{on }\iface,\label{eqn:surfaceLB_volume}%
\end{align}
where $\laplaces{}$ denotes the \name{Laplace-Beltrami} operator; cf.~subsection~\ref{subsec:laplace_beltrami_operator}. The existence and regularity of the solution $u$ can be proven by application of the according theorems of elliptic partial differential equations. At this point, it is worth noting that the regularity of $u$ crucially depends on the regularity of the underlying hypersurface $\iface$. However, since we only consider hypersurfaces of class $\mathcal{C}^\infty$ within this work, $u$ exhibits maximal regularity. Then application to the rightmost expression in \refeqn{volume_integration_base_equation} yields
\begin{align}
\int\limits_{\iface}{\iprod{\vec{x}}{\vec{n}_\iface}\,\dd{o}}=\int\limits_{\partial\iface}{\iprod{\grads{u}}{\vec{n}_{\partial\iface}}\,\dd{l}},\label{eqn:application_surface_divergence}%
\end{align}
where $\grads{}$ and $\vec{n}_{\partial\iface}$ denote the surface gradient associated to $\iface$ and the outward-pointing boundary normal, respectively. Note that $\vec{n}_{\partial\iface}$ is in the tangent space of $\iface$ at $\vec{x}_0$, i.e. $\vec{n}_{\partial\iface}\in T_\iface(\vec{x}_0)$. The introduction of appropriate boundary conditions for \refeqn{surfaceLB_volume} and properties of the sought solution are deferred to subsection~\ref{subsec:boundary_conditions}. An analytical solution to \refeqn{surfaceLB_volume} cannot be found for general hypersurfaces $\iface$. For the numerical solution within this work, we approximate the hypersurface locally and apply two different approaches: (i) a variational formulation of \refeqn{surfaceLB_volume}, using a \name{Petrov-Galerkin} approach. While the test functions are chosen to be \name{Legendre} polynomials, the choice of the ansatz functions has to be in accordance with the structure of the right-hand side, i.e.\ $\iprod{\vec{x}}{\vec{n}_\iface}$. (ii) A comparison of polynomial coefficients. The meaning and motivation for this choice will become clear below. 

\subsection{Literature review on volume computation}\label{subsec:literature_review}%
The computation of volumes emerging from the intersection of curved hypersurfaces and polygonally bounded domains (e.g., polyhedra and cuboids) has been addressed in several publications up to this date. Some of the presented approaches exploit the application of appropriate divergence theorems in order to reduce the integral dimension, while others employ direct quadrature. 

The approach of \citet{CF_2015_nioi} involves direct computation of integrals with discontinuous integrands by means of quadrature, where the boundaries of the integration domain are computed by a root finding algorithm. While their algorithm involves quite some computational effort, it is able to handle non-smooth hypersurfaces. %
\citet{JCP_2007_gioi} develop an algorithm for geometric integration over irregular domains. To obtain the hypersurface position of an intersected polyhedron, the level-set function is evaluated at the corners, allowing for a linear approximation of its respective roots on the edges. Subsequent decomposition of the polyhedron into simplices allows for straightforward evaluation of the desired integrals. %
%
\citet{JCP_2006_tnao} and the series of papers by \citet{JCP_2007_honm,JCP_2009_honm,JSC_2010_honm} are concerned with the numerical evaluation of delta-function integrals in three spatial dimensions. Considering a cuboid intersected by a hypersurface, the concept of Wen is to rewrite the integral over a three-dimensional delta-function as an integral over one of the cell faces, where the integrand is a one-dimensional delta function. All of the above approaches however imply considerable computational effort and complex case-dependent implementations.

Despite covering a different set of applications, namely the computation of integrals over implicitly given hypersurfaces, the work of \citet{IJNME_2013_hasa} is close in spirit to the present paper. The concept underlying their approach is the construction of quadrature nodes and weights from a given level-set function, where the computation of a divergence-free basis of polynomials allows to reduce the spatial problem dimension by one. By recursive application of this concept, integrals over implicitly defined domains and hypersurfaces in $\setR^3$ are transformed to line-integrals. While the method of \citet{IJNME_2013_hasa} is computationally highly efficient and exhibits high accuracy, the numerical tests shown by the authors only cover level-set functions of low polynomial order, i.e.\ hypersurfaces with few geometric details and exclusively globally convex ones. In section \ref{sec:numerical_results}, we provide results for both locally and globally non-convex hypersurfaces.
\subsection{Overall strategy}\label{subsec:overall_strategy}
The strategy of the presented algorithm consists of two parts. At first, the hypersurface $\iface$, being defined implicitly as the zero iso-contour of a level-set function $\lvlset\in\mathcal{C}^2(\cell)$, is locally represented as the graph of a (height) function $h_\iface$ over some parameter set $\paradomain_\iface\subset\setR^{d-1}$, i.e.\begin{align}
\iface=\{\vec{g}_\iface(\vec{t}):\vec{t}\in\paradomain_\iface\}\qquad\text{with}\qquad\vec{g}_\iface=\left[t_i,h_\iface(\vec{t})\right]^\mathsf{T},%
\end{align}
and parameters $\vec{t}:=\{t_i,\dots,t_{d-1}\}$. The coordinate system based in $\vec{x}_0\in\iface$ is spanned by the unit normal and the $d-1$ eigenvectors $\vec{\tau}_i$ of the associated \name{Weingarten} map, i.e.\ the directions of the principal curvatures. The associated eigenvalues are the principal curvatures $\curv_i$, corresponding to the reciprocal radii of the osculating circles. A local approximation yields a purely quadratic height function $h_\ifaceapprox$. For the remainder of this work, the approximated hypersurface will be denoted by $\ifaceapprox$, where quantities and operators introduced for $\iface$ are defined analogously. Subsection~\ref{subsec:approximation_local_coordinates} covers the mathematical details of the approximation. However, in what follows we assume the base point $\vec{x}_0$, the coordinate system $\{\vec{\tau}_i,\ns\}$ and the principal curvatures $\curv_i$ to be given. Exploiting the graph description of the interface allows to transform the integration domain to the associated parameter set $\paradomain_\iface$, i.e.
\begin{align}
\int\limits_{\iface}{\iprod{\vec{x}}{\vec{n}_\iface}\,\dd{o}}=\int\limits_{\paradomain_\iface}{\iprod{\vec{g}_\iface(\vec{t})}{\vec{n}_\iface\left(\vec{g}_\iface(\vec{\vec{t}})\right)}\fundet{\vec{g}_\iface}\,\dd{\vec{t}}},\label{eqn:integral_transformation_support}%
\end{align}
with $\fundet{\vec{g}_\iface}:=\sqrt{\det{\left(\vec{J}\!_{\vec{g}_\iface}^{\sf T}\vec{J}\!_{\vec{g}_\iface}\right)}}$ the functional determinant of $\vec{g}_\iface$, where $\vec{J}\!_{\vec{g}_\iface}$ denotes the \name{Jacobian}. To facilitate the numerical treatment, the parameter set is approximated by a polygon which, in general, is neither a super- nor a subset of the true parameter set, cf.~\reffig{support_triangulation}. We will discuss the implications of this property in subsection~\ref{subsec:petrov_galerkin}. The second part of the strategy is a numerical solution of the surface \name{Laplace-Beltrami} equation. The first concept comprises the application of a \name{Petrov-Galerkin} approach on the variational formulation, i.e.\ $\laplaces[\ifaceapprox]{u}=\iprod{\vec{x}}{\vec{n}_\ifaceapprox}$ is replaced by
\begin{align}
\sum\limits_{j=1}^{N}{\hat{u}_j\int\limits_{\ifaceapprox}{\testfun[i]\laplaces[\ifaceapprox]{\ansatzfun[j]}}\,\dd{o}}=\int\limits_{\ifaceapprox}{\iprod{\vec{x}}{\vec{n}_\ifaceapprox}\testfun[i]\,\dd{o}}\qquad\forall\quad 1\leq i\leq N\qquad\text{with}\qquad u=\sum\limits_{j=1}^{N}{\hat{u}_j\ansatzfun[j]},\label{eqn:surfaceLB_variational_formulation_introduction}%
\end{align}
where $\testfun[i]$ and $\ansatzfun[j]$ are the test and ansatz functions, respectively. The derivation of \refeqn{surfaceLB_variational_formulation_introduction} along with a sketch of the solution strategy are the subject of subsection~\ref{subsec:petrov_galerkin}. The second concept involves the comparison of coefficients of a polynomial expression, allowing to restrict the deviation of the exact and numerical solution to polynomials of higher order, which become negligible for sufficiently small parameter sets. Subsection~\ref{subsec:equating_polynomial_coefficients} provides the details.
%
%
\subsection{Outline}\label{subsec:outline}
Section \ref{sec:mathematical_details} introduces the notation and derives mathematical details for two and three spatial dimensions, where basic facts from differential geometry are placed to \ref{app:review_differential_geometry}. Since the representation of hypersurfaces is of key importance, subsection \ref{subsec:approximation_local_coordinates} comprises the introduction of local coordinates, as well as an approximation using the \name{Weingarten} map. Subsequently, we introduce the \name{Laplace-Beltrami} operator, both in local coordinates and in a comprehensible level-set notation. For certain classes of hypersurfaces, \refeqn{surfaceLB_volume} admits analytical solutions, which will be presented and employed to discuss the admissibility of boundary conditions for \refeqn{surfaceLB_volume}. Finally, this section comprises the numerical solution approaches, namely the comparison of coefficients of polynomials as well as the \name{Petrov-Galerkin} approach, with a focus on the parameter set $\paradomain_\ifaceapprox$ of the graph representation of $\ifaceapprox$. %
Section \ref{sec:numerical_algorithm} introduces the numerical algorithm, where details of the implementation are provided both for the approximation of the hypersurface and the assembly and solution of the linear system of equations resulting from the variational formulation. Moreover, we provide some details of the coefficient comparison. %
Section \ref{sec:numerical_results} is concerned with several numerical experiments for $d=3$ spatial dimensions and discusses the results. Finally, section \ref{sec:conclusion} concludes the presented work and formulates a further outlook. 
\section{Mathematical concept of the approach}\label{sec:mathematical_details}%
%
%
\subsection{Boundary conditions}\label{subsec:boundary_conditions}%
Note that the application of the divergence theorem, cf.~\refeqn{application_surface_divergence}, does not require any boundary conditions for the sought function $u$. In order to facilitate numerical treatment by exploitation of divergence theorems, it is favorable to either prescribe \name{Dirichlet} or \name{Neumann} boundary conditions. While in theory, the problem at hand does admit solutions\footnote{For further mathematical details on the existence of solutions, the reader is referred to \citet{Pruess2016} and the references given therein.} fulfilling \name{Dirichlet} conditions, say, e.g., $u_{\rvert\partial\iface}=0$, the desired application of the surface divergence theorem, cf.~\refeqn{application_surface_divergence}, obviously prohibits homogeneous \name{Neumann} boundary conditions, because $\int_{\partial\iface}{\iprod{\grads{u}}{\vec{n}_{\partial\iface}}\dd{l}}=0$ for $\grads{u}_{\rvert\partial\iface}=\vec{0}$. In the context of the numerical algorithm presented here, however, we are only interested in the approximation of \textit{any} regular solution $u$, whose surface gradient is evaluated on $\partial\iface$. With an appropriate ansatz space ensuring regularity, both the variational formulation and the comparison of coefficients provide a unique solution. Hence, the presented approach does not require to specify particular boundary conditions. The admissibility of \name{Dirichlet} boundary conditions is deferred to the last paragraph in subsection~\ref{subsec:laplace_beltrami_operator}. 
\subsection{Approximation of hypersurfaces in local coordinates}\label{subsec:approximation_local_coordinates}%
As shown in the appendix, cf.~\refeqn{surface_parametrization}, under the general assumptions formulated above, the hypersurface patch $\ifaceapprox$ can be parametrized as the graph of a height function, i.e.\footnote{To ease notation and avoid explicit notations for different values of $d$, henceforth the \name{Einstein} summation convention applies.}
\begin{align}
\iface=\{\vec{x}_0+t_i\vec{\tau}_i+h_\iface(\vec{t})\vec{n}_0:\vec{t}\in\paradomain_\iface\}\quad\text{with}\quad h_\iface\in\mathcal{C}^2\left(\paradomain_\iface\right)\quad\text{and}\quad\vec{n}_0:=\vec{n}_\iface(\vec{x}_0),%
\end{align}
where $\{\vec{\tau}_i,\vec{n}_0\}$ forms an orthonormal system for fixed $\vec{x}_0$. Also, an appropriate shift of coordinates ensures $\vec{x}_0=\vec{0}$. The computation of the height function $h_\iface=h_\iface(\vec{t};\vec{x}_0)$ requires to solve the nonlinear implicit equation $\lvlset_\iface(\vec{x}_0+t_i\vec{\tau}_i+h_\iface\vec{n}_0)=0$. Since this may be cumbersome, we choose to approximate the hypersurface around $\vec{x}_0$ by the graph of an approximated height function, based on the principal curvatures provided by the \name{Weingarten} map. We obtain
\begin{align}
\ifaceapprox=\{\vec{g}_{\ifaceapprox}(\vec{t}):\vec{t}\in\paradomain_{\ifaceapprox}\}\quad\text{with}\quad\vec{g}_{\ifaceapprox}(\vec{t}):=\vec{x}_0+t_i\vec{\tau}_i+h_{\ifaceapprox}\vec{n}_0,\label{eqn:approx_hypersurface_graph_representation}%
\end{align}
where the height function reads
\begin{align}
h_{\ifaceapprox}=\frac{1}{2}\sum\limits_{i=1}^{d-1}{\curv_it_i^2}, \qquad\text{with}\qquad\norm{h_\iface-h_{\ifaceapprox}}=\mathcal{O}\left(\lVert\vec{t}\rVert^4\right).\label{eqn:hypersurface_height_approx_hot}%
\end{align}
Note that in general, as mentioned above, the respective parameter sets   do \textit{not} coincide, i.e.\ $\paradomain_\iface\neq\paradomain_{\ifaceapprox}$. However, the parameter set deviation $\Delta\paradomain_\iface:=(\paradomain_{\iface}\setminus\paradomain_{\ifaceapprox})\cup(\paradomain_{\ifaceapprox}\setminus\paradomain_\iface)$ will be small if the characteristic length $d_\cell$ of cell $\cell$ suffices $d_\cell\curv_i\leq\num{e-2}$, see~\reffig{support_deviation_approxiamtion} for an illustration.

\begin{figure}[h]
\null\hfill%
\subfigure[hypersurface patches]{\includegraphics{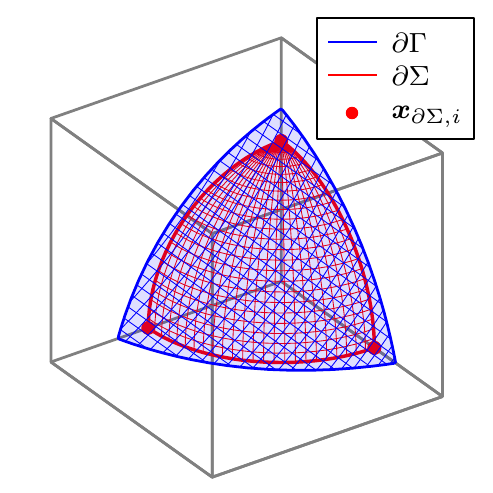}\label{fig:support_deviation_approxiamtion_01}}%
\hfill%
\subfigure[associated parameter sets  ]{\includegraphics{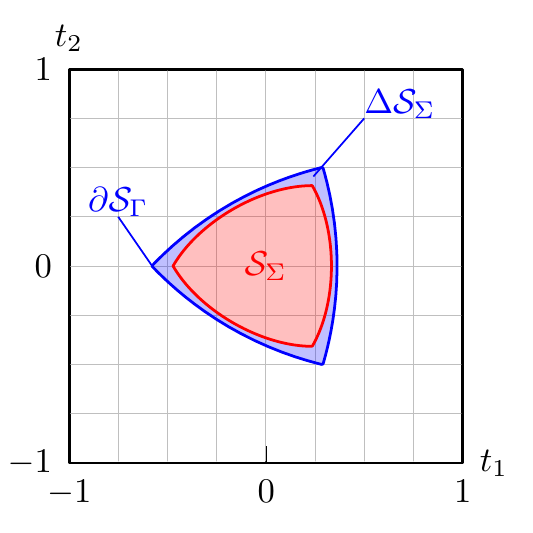}\label{fig:support_deviation_approxiamtion_02}}%
\hfill\null%
\caption{Deviation of parameter sets induced by approximation (light blue) of the hypersurface $\iface=\partial\mathcal{B}_R(\vec{0})\cap\cell$ (red mesh) around $\vec{x}_0=\frac{1}{3}{[1,1,1]}$ with $\curv_i=\frac{1}{R}$ and $\vec{n}_0={\frac{-1}{\sqrt{3}}[1,1,1]}$. In general, it holds that $\paradomain_{\ifaceapprox}\not\supset\paradomain_{\iface}$, i.e.\ the parameter set of the approximation does not contain the true parameter set.}%
\label{fig:support_deviation_approxiamtion}%
\end{figure}
Furthermore, the graph representation of $\iface$, cf.~\refeqn{approx_hypersurface_graph_representation}, allows to assign to any $f:\iface\mapsto\setR$ a function $f^\iface:\paradomain_\iface\mapsto\setR$ with $f^\iface:=f\circ\vec{g}_\iface$.

%
%
\subsection{Representations of the Laplace-Beltrami operator}\label{subsec:laplace_beltrami_operator}
On a curved manifold $\iface$ the correspondent to the \name{Laplace} operator $\laplace{u}=\partial_{ii}u$ in \name{Euclidean} space, being defined as the divergence of the gradient of a scalar function is the \name{Laplace-Beltrami} operator (associated to $\iface$), defined as $\laplaces{u}:=\divs{\grads{u}}$. In what follows, we derive the concrete form of the \name{Laplace-Beltrami} operator for implicitly (in terms of a level-set) and explicitly (as the graph of a function) defined hypersurfaces. For the level-set case, the authors could not find the specific representations in the literature. Furthermore, we present specific analytical solutions of the \name{Laplace-Beltrami} equation emerging from the computation of volumes, cf.~\refeqn{surfaceLB_volume}. In the sequel, $\ifaceapprox$ represents a member of the class of hypersurfaces given by \refeqn{approx_hypersurface_graph_representation}.
\paragraph{Level-set representation}
For a hypersurface $\iface\subset\setR^d$ defined by the iso-contour of a smooth level-set $\lvlset\in\mathcal{C}^2(\setR^d)$, one obtains 
\begin{align}
\laplaces{u}&=%
\PS:\hessian{u}-\frac{\iprod{\grad{u}}{\grad{\lvlset}}}{\iprod{\grad{\lvlset}}{\grad{\lvlset}}}\left(\PS:\hessian{\lvlset}\right)%
=\PS:\hessian{u}-\iprod{\grad{u}}{\ns}\frac{\PS:\hessian{\lvlset}}{\iprod{\grad{\lvlset}}{\grad{\lvlset}}^{\frac{1}{2}}}
,\label{eqn:surfaceLB_definition_levelset}%
\end{align}
where $\PS:=\ident-\dyad{\ns}{\ns}$, $\hessian{f}:=\partial_{ij}f\vec{e}_i\otimes\vec{e}_j$ and $\vec{A}:\vec{B}:=\trace{\left(\vec{A}^{\mathsf{T}}\vec{B}\right)}$ denote the  tangential projection, \name{Hessian} matrix and real tensor contraction, respectively. 
Here, we would like to emphasize the relation to the mean curvature
\begin{align}
\curv_\iface:=\divs{\left(-\ns\right)}=\sum_{i=1}^{d-1}\curv_i=-\frac{\PS:\hessian{\lvlset}}{\iprod{\grad{\lvlset}}{\grad{\lvlset}}^{\frac{1}{2}}}.\label{eqn:mean_curvature_definition}%
\end{align}
\paragraph{Graph representation ($\bf{d=2}$)}%
If the hypersurface is given as the graph of a function $h_\iface(\vec{t}):\paradomain_\iface\mapsto\setR$ with parameter set $\paradomain_\iface\subset\setR$, 
introducing $\partial_i:=\frac{\partial}{\partial t_i}$ and $\partial_{ij}:=\frac{\partial^2}{\partial t_i\partial t_j}$, one obtains
\begin{align}
\laplaces{u}=\partial_{11}u\frac{1}{1+\left(\partial_1h_\iface\right)^2}-\partial_1u\frac{\partial_{11}h_\iface\partial_1h_\iface}{\left(1+\left(\partial_1h_\iface\right)^2\right)^2}.\label{eqn:surfaceLB_definition_graph_2D}%
\end{align}
The above form is easily derived from \refeqn{surfaceLB_definition_levelset} with $\lvlset(\vec{t})=t_2-h_\iface(t_1)$ and $u(\vec{t})=u(t_1)$.  For height functions of purely quadratic form, i.e.\ $h_\ifaceapprox=\frac{\curv t_1^2}{2}$,  cf.~\refeqn{approx_hypersurface_graph_representation}, \refeqn{surfaceLB_definition_graph_2D} becomes
\begin{align}
\laplaces[\ifaceapprox]{u}=\partial_{11}u\frac{1}{1+\curv^2t_1^2}-\partial_1u\frac{\curv^2t_1}{\left(1+\curv^2t_1^2\right)^2}.\label{eqn:surfaceLB_definition_approx_graph_2D}%
\end{align}
The right-hand sides become
\begin{align}
\iprod{\vec{x}}{\vec{n}_\iface}=\frac{-h_\iface}{\sqrt{1+\left(\partial_1h_\iface\right)^2}}\qquad\text{and}\qquad\iprod{\vec{x}}{\vec{n}_\ifaceapprox}=\frac{-\curv_1t_1^2}{2\sqrt{1+\curv_1^2t_1^2}}.\label{eqn:surfaceLB_graph_rhs_2D}%
\end{align}
\paragraph{Graph representation ($\bf{d=3}$)}%
By arguments analogous to those given above, for the case of three spatial dimensions we have $\lvlset(\vec{t})=t_3-h_\iface(t_1,t_2)$, yielding
\begin{align}
\laplaces{u}&=%
\partial_{11}u\frac{1+\partial_2h_\iface^2}{1+\partial_1h_\iface^2+\partial_2h_\iface^2}+\partial_{22}u\frac{1+\partial_1h_\iface^2}{1+\partial_1h_\iface^2+\partial_2h_\iface^2}-\partial_{12}u\frac{2\partial_1h_\iface\partial_2h_\iface}{1+\partial_1h_\iface^2+\partial_2h_\iface^2}\nonumber\\%
&-\frac{\partial_1u\partial_1h_\iface+\partial_2u\partial_2h_\iface}{\left(1+\partial_1h_\iface^2+\partial_2h_\iface^2\right)^2}\left(\partial_{11}h_\iface\left(1+\partial_2h_\iface^2\right)+\partial_{22}h_\iface\left(1+\partial_1h_\iface^2\right)-2\partial_{12}h_\iface\partial_1h_\iface\partial_2h_\iface\right),\label{eqn:surfaceLB_definition_graph_3D}%
\end{align}
which in the purely quadratic case, i.e.\ with $h_\ifaceapprox=\frac{\curv_1t_1^2+\curv_2t_2^2}{2}$, simplifies to
\begin{align}
\laplaces[\ifaceapprox]{u}&=%
\partial_{11}u\frac{1+\curv_2^2t_2^2}{1+\curv_1^2t_1^2+\curv_2^2t_2^2}+\partial_{22}u\frac{1+\curv_1^2t_1^2}{1+\curv_1^2t_1^2+\curv_2^2t_2^2}-\partial_{12}u\frac{2\curv_1\curv_2t_1t_2}{1+\curv_1^2t_1^2+\curv_2^2t_2^2}\nonumber\\%
&-\frac{\partial_1u\curv_1t_1+\partial_2u\curv_2t_2}{\left(1+\curv_1^2t_1^2+\curv_2^2t_2^2\right)^2}\left(\curv_1\left(1+\curv_2t_2^2\right)+\curv_2\left(1+\curv_1t_1^2\right)\right).\label{eqn:surfaceLB_definition_approx_graph_3D}%
\end{align}
Note that the application of the operator given in \refeqn{surfaceLB_definition_approx_graph_3D} to a function preserves the following symmetry:
\begin{align}
\begin{split}
u(t_1,t_2)=u(t_1,-t_2)\implies[\laplaces{u}(\vec{t})]_{\vert\vec{t}=[x_1,x_2]}=[\laplaces{u}(\vec{t})]_{\vert\vec{t}=[x_1,-x_2]}\\%
u(t_1,t_2)=u(-t_1,t_2)\implies[\laplaces{u}(\vec{t})]_{\vert\vec{t}=[x_1,x_2]}=[\laplaces{u}(\vec{t})]_{\vert\vec{t}=[-x_1,x_2]}
\end{split}
.\label{eqn:surfaceLB_symmetry_preservation}%
\end{align}
Analogously to the case above, the right-hand sides become
\begin{align}
\iprod{\vec{x}}{\vec{n}_\iface}=\frac{-h_\iface}{\sqrt{1+\left(\partial_1h_\iface\right)^2+\left(\partial_2h_\iface\right)^2}}\qquad\text{and}\qquad\iprod{\vec{x}}{\vec{n}_\ifaceapprox}=\frac{-\curv_1t_1^2-\curv_2t_2^2}{2\sqrt{1+\curv_1^2t_1^2+\curv_2^2t_2^2}}.\label{eqn:surfaceLB_graph_rhs_3D}%
\end{align}
\paragraph{Analytical solutions for ${\laplaces[\ifaceapprox]{u}}=\iprod{\vec{x}}{\vec{n}_\ifaceapprox}$}%
For non-planar hypersurfaces $\ifaceapprox$ in two spatial dimensions with arbitrary but constant $\curv\in\setR$, combining \refeqn{surfaceLB_definition_approx_graph_2D} and \refeqn{surfaceLB_graph_rhs_2D}, we obtain the family of solutions 
\begin{align}
u(t_1;\curv)=u_0+\left(\frac{1}{18\curv^3}-\frac{1+\curv^2t_1^2}{30\curv^3}\right)\left(1+\curv^2t_1^2\right)^{\frac{3}{2}},\label{eqn:surfaceLB_approx_solution_2D}%
\end{align}
where choosing $u_0=-\frac{1}{45\curv^3}$ ensures $u(0;\curv)=0$. For planar hypersurfaces, the solution becomes trivial, since $\lim_{\curv\to0}u(t_1;\curv)=0$. The existence of an analytical solution implies that the computation of the volume (i.e.\ the area, since we consider $d=2$ here) only requires to compute the two intersections, denoted $t_1^\pm$, of the approximated hypersurface $\ifaceapprox$ with the cell boundary $\partial\cell$, e.g. by a simple \name{Newton} algorithm. The $t_1^\pm$ are plugged then into \refeqn{surfaceLB_approx_solution_2D} to obtain the surface gradient $\grads{u}\rvert_{t_1=t_1^\pm}$, which is then used to evaluate the inner product with the boundary normal $\vec{n}_{\partial\iface}$. \refFig{surfaceLB_evaluation_example_2D} illustrates the relevant quantities. Also, an advantage of our approach becomes evident in \reffig{surfaceLB_evaluation_example_2D_01}: for $d=2$, the approximated hypersurface is not required to be the graph of a function whose independent variable varies along one of the cell edges.    

\begin{figure}[h]
\null\hfill%
\subfigure[hypersurface patch with boundary normals]{\includegraphics{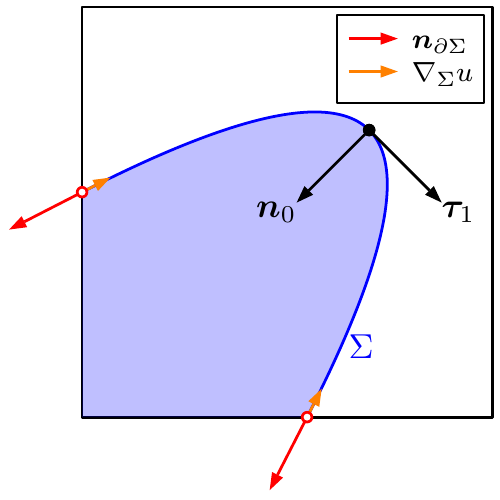}\label{fig:surfaceLB_evaluation_example_2D_01}}%
\hfill
\subfigure[analytical solution, cf.~\refeqn{surfaceLB_approx_solution_2D}]{\includegraphics{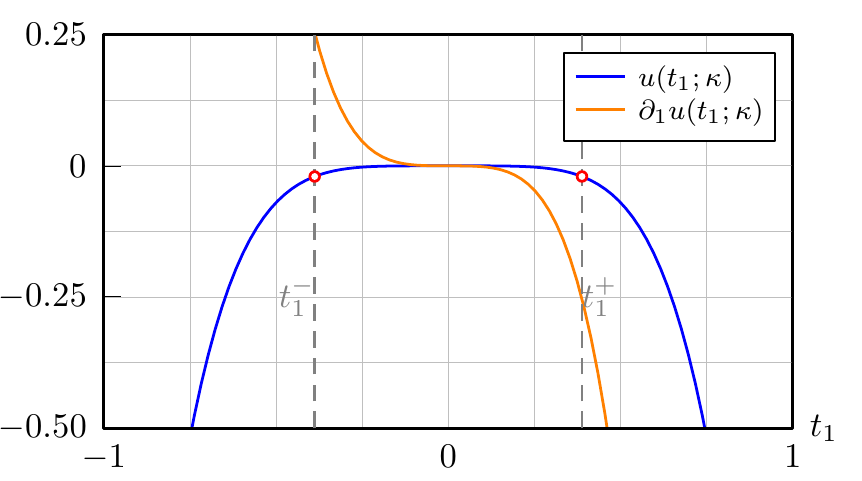}\label{fig:surfaceLB_evaluation_example_2D_02}}%
\hfill\null%
\caption{$\curv=8$, $\vec{n}_0=\frac{1}{\sqrt{2}}{[-1,-1]^{\mathsf{T}}}$, $\vec{x}_0=\frac{7}{10}{[1,1]^{\mathsf{T}}}$}%
\label{fig:surfaceLB_evaluation_example_2D}
\end{figure}

In three spatial dimensions, cf.~\refeqn{surfaceLB_definition_approx_graph_3D} and \refeqn{surfaceLB_graph_rhs_3D}, a family of analytical solutions can be given for coinciding and constant principal curvatures $\curv_1=\curv_2=\curv$, yielding
\begin{align}
u(\vec{t};\curv)=u_0+\left(\frac{1}{24\curv^3}-\frac{1+\curv^2t_1^2+\curv^2t_2^2}{40\curv^3}\right)\left(1+\curv^2t_1^2+\curv^2t_2^2\right)^{\frac{3}{2}}.\label{eqn:surfaceLB_approx_solution_3D}%
\end{align}
By choosing $u_0=-\frac{1}{60\curv^3}$, one obtains $u(\vec{0};\curv)=0$. For the non-trivial case $\curv\neq0$, the iso-contours of the analytical solution, i.e.\ {$\mathcal{I}(u;\alpha):=\{\vec{t}\in\setR^2:u(\vec{t};\curv)=\alpha\}$}, are circles. This implies that for $\alpha\neq0$ on a polygonal parameter set $\paradomain_\ifaceapprox$, which is preferable for numerical implementation, the function $u_{\rvert\partial\paradomain_\ifaceapprox}$ cannot be constant, especially $u\rvert_{\partial\paradomain_\ifaceapprox}\neq0$; cf.~\reffig{surfaceLB_analytical_solution_3D}. This imposes crucial restrictions on the numerical algorithm for the solution of the variational problem, if, e.g., one seeks to exploit partial integration; cf.~\refeqn{trafo_variational_formulation}.

\begin{figure}[h]
\null\hfill%
\includegraphics{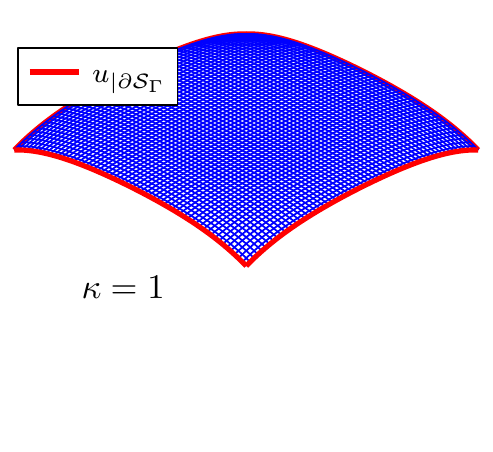}%
\hfill
\includegraphics{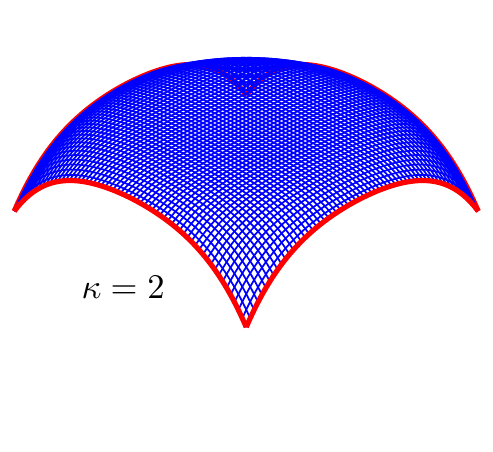}%
\hfill
\includegraphics{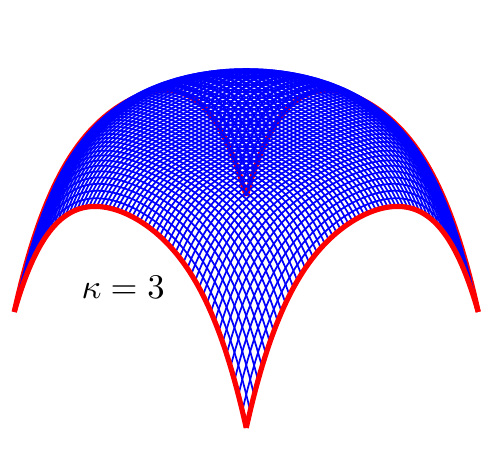}%
\hfill\null%
\caption{Visualization of \refeqn{surfaceLB_approx_solution_3D} for $\paradomain_\ifaceapprox={[-1,1]^2}$ and various $\curv$.}%
\label{fig:surfaceLB_analytical_solution_3D}
\end{figure}

For the general case $\curv_1\neq\curv_2$, an analytical solution could not be found by the authors. Hence, we transform the problem into its variational formulation, in order to make it accessible for numerical treatment.
%
%
\subsection{Variational formulation and Petrov-Galerkin ansatz}\label{subsec:petrov_galerkin}%
%
%
The present subsection is devoted to the variational formulation of $\laplaces[\ifaceapprox]{u}=\iprod{\vec{x}}{\vec{n}_\ifaceapprox}$ for hypersurfaces $\ifaceapprox\subset\setR^3$ defined by \refeqn{approx_hypersurface_graph_representation}, following a standard approach: we multiply by a test function $\testfun[i]\in\testfunset$, approximate the sought solution $u$ by a series of $N$ ansatz functions $\ansatzfun[j]\in\ansatzfunset$ and numerically integrate over $\ifaceapprox$. Note that because the properties of the analytical solution given in \refeqn{surfaceLB_approx_solution_3D}, which is desired to be an element of the ansatz function space, prohibit application of \name{Dirichlet} boundary conditions on polygonally bounded parameter sets, and \name{Neumann} boundary conditions are incompatible within our approach, we do not apply partial integration. The details of the function spaces are provided below. As stated in subsection~\ref{subsec:approximation_local_coordinates}, due to the explicit parametrization, any function  $f$ mapping from the hypersurface $\ifaceapprox$ may be expressed as $f^\ifaceapprox=f\circ\vec{g}_\ifaceapprox$, with $f^\ifaceapprox:\paradomain_\ifaceapprox\mapsto\setR$. Exploiting the integral transformation from \refeqn{integral_transformation_support}, one obtains
\begin{align}
\sum\limits_{j=1}^{N}{\hat{u}_j\int\limits_{\paradomain_\ifaceapprox}{\testfun[i](\vec{t})\laplaces[\ifaceapprox]{\ansatzfun[j](\vec{t})}}\fundet{\vec{g}_\ifaceapprox(\vec{t})}\dd{\vec{t}}}=\int\limits_{\paradomain_\ifaceapprox}{f^\ifaceapprox\!(\vec{t})\testfun[i](\vec{t})\fundet{\vec{g}_\ifaceapprox(\vec{t})}\dd{\vec{t}}}\qquad\text{or}\qquad\vec{A}_\ifaceapprox\hat{\vec{u}}=\vec{b}_\ifaceapprox,\label{eqn:surfaceLB_variational_formulation}%
\end{align}
with the functional determinant $\fundet{\vec{g}_\ifaceapprox}:=\iprod{\grad{\lvlset_\ifaceapprox}}{\grad{\lvlset_\ifaceapprox}}^{\frac{1}{2}}=\sqrt{1+\curv_1^2t_1^2+\curv_2^2t_2^2}$ corresponding to the area of an infinitesimal hypersurface element; cf.~subsection \ref{subsec:laplace_beltrami_operator}. 
\paragraph{Approximation of the parameter set:}%
A direct numerical quadrature of \refeqn{surfaceLB_variational_formulation} is difficult due to the potentially non-polygonal shape of the parameter set $\paradomain_\ifaceapprox$. Therefore we approximate the parameter set by a polygon spanned by the projection of the intersections of the hypersurface with the cell edges; cf.~\reffig{support_triangulation_02}. 

\begin{figure}[h]
\null\hfill%
\subfigure[hypersurface patch]{\includegraphics{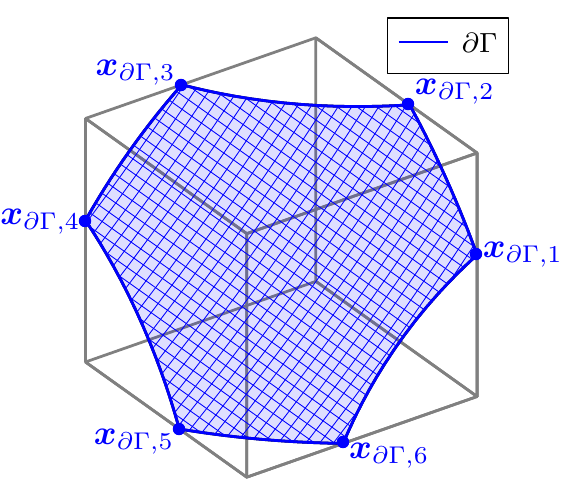}\label{fig:support_triangulation_01}}%
\hfill%
\subfigure[true and triangulated parameter set]{\includegraphics{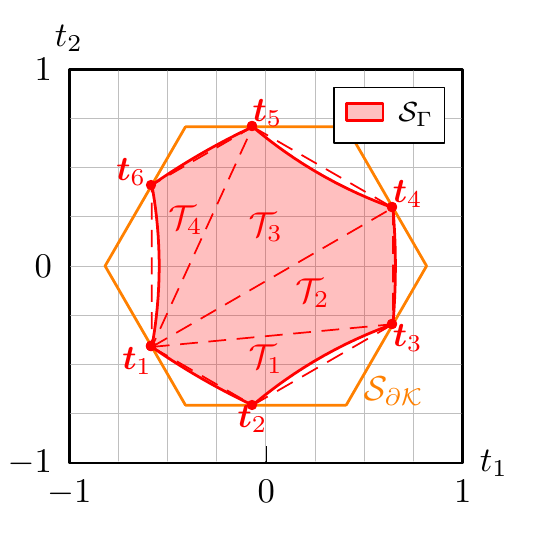}\label{fig:support_triangulation_02}}%
\hfill\null%
\caption{Hypersurface ($\vec{x}_0=\frac{6}{10}{[1,1,1]}$, $\curv_i=-\frac{1}{2}$ and $\vec{n}_0={\frac{-1}{\sqrt{3}}[1,1,1]}$) with true (shaded) and triangulated (dashed lines) parameter set, where $\paradomain_{\partial\cell}$ is the projection of the cell boundary. Note that the polygon spanned by the projections of the edge intersections (\textcolor{red}{\textbullet}) does not contain the true parameter set, nor vice versa, i.e.\ $\bigcup_k\triangle_k\not\supset\paradomain_\ifaceapprox$ and $\bigcup_k\triangle_k\not\subset\paradomain_\ifaceapprox$, in general.}%
\label{fig:support_triangulation}%
\end{figure}

The integration over the approximated parameter set $\paradomain_\ifaceapprox\approx\bigcup_{k}\triangle_k$ can then be performed by transformation of the respective triangles to the referential square $\mathcal{S}_0:=[0,1]^2$ (via the referential triangle $\triangle_0$) and standard \name{Gauss-Legendre} quadrature, i.e.
\begin{align}
\int\limits_{\triangle_k}{f(\vec{t})\,\dd{\vec{t}}}=\int\limits_{\triangle_0}{f\!\left(\vec{T}_k(\vec{u})\right)\lVert\det\vec{J}_{\vec{T}_k}\rVert\,\dd{\vec{u}}}=\int\limits_{\mathcal{S}_0}{f\!\left(\vec{T}_k\left(\vec{T}_{\mathcal{S}}(\vec{u})\right)\right)u_1\lVert\det\vec{J}_{\vec{T}_k}\rVert\,\dd{\vec{u}}}\approx\sum\limits_{i}{f(\hat{\vec{t}}_{k,i})\omega^{\ifaceapprox}_{k,i}},\label{eqn:triangle_transformation_quadrature}%
\end{align}
where $\hat{\vec{t}}_{i,k}\in\triangle_k$ are the quadrature nodes with associated weights $\omega^{\ifaceapprox}_{k,i}$; cf.~\reffig{triangle_transformation} for an illustration.

\begin{figure}[h]
\centering
\includegraphics{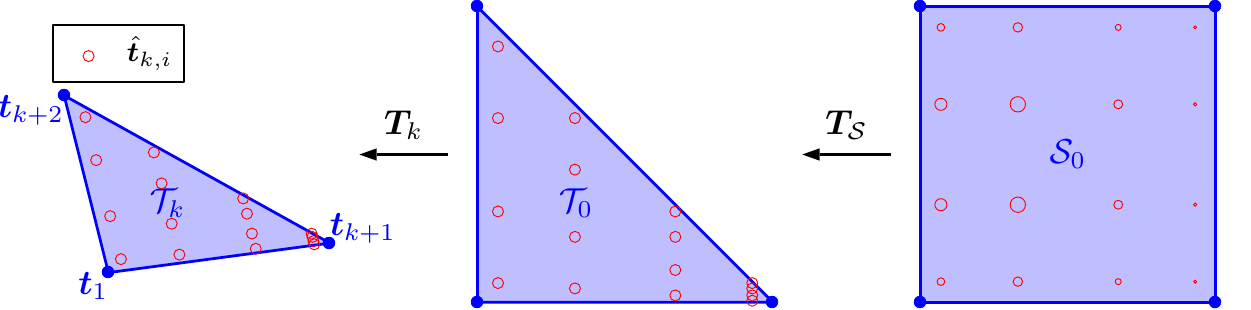}%
\caption{Transformation of triangular integration domain $\triangle_k\subset\setR^2$ to $\mathcal{S}_0$. In the reference square, the circle sizes indicate the relative magnitude of the weights $\omega^{\ifaceapprox}_{k,i}$.}%
\label{fig:triangle_transformation}%
\end{figure}

Although it would be more convenient to approximate the parameter set by the projection of the cell boundary $\paradomain_{\partial\cell}$, say, our numerical experiments have shown that the quality of the parameter set approximation is crucial for the overall accuracy of the algorithm. The accuracy especially suffers from a coarse parameter set approximation for $\curv_1\neq\curv_2$, rapidly decreasing for increasing $\lvert\lvert\curv_1\rvert-\lvert\curv_2\rvert\rvert$. 

\paragraph{The choice of ansatz and test functions:}%
In order for \refeqn{surfaceLB_variational_formulation} to be a well-posed problem for $\vec{\hat{u}}$, the regularity required for the test functions is $\testfun[i]\in\testfunset\subset\squareInt{\paradomain_\ifaceapprox}$ and for the ansatz functions $\ansatzfun[j]\in\ansatzfunset\subset\hilbert{\paradomain_\ifaceapprox}$, respectively. Since the test functions need not contain any information on the underlying hypersurface, let $\testfunset=\mathcal{L}_N(t_1)\times\mathcal{L}_N(t_2)$, where $\mathcal{L}_N(x):=\{P_k(x):0\leq k\leq N\}$ is the set of \name{Legendre} polynomials up to order $N$. Preliminary numerical experiments indicated that the ansatz functions $\ansatzfun[i]$ have to be chosen in accordance with both the underlying hypersurface $\ifaceapprox$ and the right-hand side. Here, the ansatz functions contain the norm of an infinitesimal surface element, i.e.
\begin{align}
\ansatzfun[i]:=\testfun[i]\iprod{\grad{\lvlset_\ifaceapprox}}{\grad{\lvlset_\ifaceapprox}}^{\frac{3}{2}}=\testfun[i]\left(1+\curv_1^2t_1^2+\curv_2^2t_2^2\right)^{\frac{3}{2}}.\label{eqn:ansatz_function_definition}%
\end{align}
This choice also ensures that the analytical solution for the volume computation case ($f=\iprod{\vec{x}}{\vec{n}_\ifaceapprox}$) given in \refeqn{surfaceLB_approx_solution_3D} is an element of $\operatorname{span}(\ansatzfunset)$, which is not possible by choosing polynomial ansatz functions. Also, we would like to emphasize that the above mentioned symmetry properties of the \name{Laplace-Beltrami} operator, cf.~\refeqn{surfaceLB_symmetry_preservation}, for purely quadratic hypersurfaces allows to remove those \name{Legendre} polynomials with odd order, since their contributions cancel during the integration.  
%
%
\subsection{Equating polynomial coefficients}\label{subsec:equating_polynomial_coefficients}%
Within this subsection, let $g:=1+\curv_1^2t_1^2+\curv_2^2t_2^2$ for ease of notation and assume that $u:\setR^2\mapsto\setR$ is polynomial. Note that for the gradient and \name{Hessian} matrix, respectively, one obtains
\begin{align}
\partial_i\!\left(u\sqrt{g^3}\right)&=\frac{1}{\sqrt{g}}\left[g^2\partial_if+\frac{3}{2}fg\partial_ig\right],\nonumber\\%
\partial_{ij}\!\left(u\sqrt{g^3}\right)&=\frac{1}{\sqrt{g}}\left[g^2\partial_{ij}u+\frac{3g}{2}\left(\partial_ig\partial_jf+\partial_jg\partial_if+u\partial_{ij}g\right)+\frac{3f}{2}\partial_ig\partial_jg\right],%
\end{align}
where the expressions in parentheses are also polynomial. Inserting the above into the definition of the surface \name{Laplace-Beltrami} operator, cf.~\refeqn{surfaceLB_definition_approx_graph_3D}, and comparing the result with the right-hand side, cf.~\refeqn{surfaceLB_graph_rhs_3D}, it becomes evident that the left-hand side of 
\begin{align}
\sqrt{g}\left(\laplaces[\ifaceapprox]{\left(u\sqrt{g^3}\right)}-\iprod{\vec{x}}{\vec{n}_\ifaceapprox}\right)=0\label{eqn:surfaceLB_polynomial_comparison_base}%
\end{align}
is a polynomial expression. In fact, one obtains
\begin{align}
\sqrt{g}\laplaces[\ifaceapprox]{\left(t_1^{m}t_2^{n}\sqrt{g^3}\right)}=&\phantom{+}(\curv_1^2(m^2+4m+3) + (n^2+4n+3)\curv_2^2 - \curv_1\curv_2(2mn+m+n))t_1^mt_2^n\nonumber\\%
&+(\curv_1^2\curv_2^2(m^2+4m+3) - \curv_1\curv_2^3(2mn+6m+n+3))t_1^mt_2^{n+2} \nonumber\\%
&+(\curv_1^2\curv_2^2(n^2+4n+3) - \curv_1^3\curv_2(2mn+6n+m+3))t_1^{m+2}t_2^n \nonumber\\%
&+2\curv_2^2(m^2 - m)t_1^{m-2}t_2^{n+2}+(n^2-n)t_1^mt_2^{n-2}+(m^2-m)t_1^{m-2}t_2^n  \nonumber\\%
&+\curv_2^4(m^2-m)t_1^{m-2}t_2^{n+4} +2\curv_1^2(n^2-n)t_1^{m+2}t_2^{n-2} \nonumber\\%
&+\curv_1^4(n^2-n)t_1^{m+4}t_2^{n-2}.\label{eqn:surfaceLB_polynomial_mapping}%
\end{align}
Furthermore, the symmetry of $\iprod{\vec{x}}{\vec{n}_\ifaceapprox}$ implies that any solution $u$ of \refeqn{surfaceLB_polynomial_comparison_base} can only contain even powers of $t_i$, hence we choose the ansatz
\begin{align}
u=\sum\limits_{i=0}^{N}{\sum\limits_{j=0}^{M}{\hat{u}_{ij}t_1^{2i}t_2^{2j}}},%
\end{align}
where the $(N+1)(M+1)$ coefficients $\vec{\hat{u}}:=\{\hat{u}_{ij}\}$ are obtained from comparison of polynomial coefficients. As can be seen from \refeqn{surfaceLB_polynomial_mapping}, the modified \name{Laplace-Beltrami} operator $\tilde{\Delta}_\ifaceapprox{u}:=\sqrt{g}\laplaces[\ifaceapprox]{(u\sqrt{g^3})}$ expands the polynomial span of its argument, implying that the system of equations governing the coefficients $\hat{u}_{ij}$ will be overdetermined for general $\curv_1\neq\curv_2$, since
\begin{align}
\sum\limits_{i=0}^{N}{\sum\limits_{j=0}^{M}{\hat{u}_{ij}\tilde{\Delta}_\ifaceapprox{(t_1^{2i}t_2^{2j})}}}-\sqrt{g}\iprod{\vec{x}}{\vec{n}_\ifaceapprox}=\sum\limits_{i=0}^{N+2}{\sum\limits_{j=0}^{M+2}{\beta_{ij}(\vec{\hat{u}};\curv_1,\curv_2)\,t_1^{2i}t_2^{2j}}}.%
\end{align}
Solving \refeqn{surfaceLB_polynomial_comparison_base} exactly is equivalent to finding $\vec{\hat{u}}$ such that $\vec{\beta}(\vec{\hat{u}};\curv_1,\curv_2)=\vec{0}$. Since $\vec{\beta}$ is linear in $\vec{\hat{u}}$, we may write
\begin{align}
\vec{B}_\ifaceapprox\vec{\hat{u}}=\vec{q}_\ifaceapprox\qquad\text{with}\qquad \vec{B}_\ifaceapprox\in\setR^{K\times(N+1)(M+1)}\quad\text{and}\quad\vec{q}_\ifaceapprox\in\setR^K,\label{eqn:polynomial_comparison_linear_system}%
\end{align}
where the number of rows $K$ is a function of maximum polynomial oders $N,M$ with $K\geq(N+1)(M+1)$. Numerical experiments for $1\leq N,M\leq6$ indicate that (i) the matrix $\vec{B}_\ifaceapprox$ does not have full rank, i.e.\ $\operatorname{rank}(\vec{B}_\ifaceapprox)<K$, and (ii) the rank of $\vec{B}_\ifaceapprox$ is $(N+1)(M+1)$, cf.~\reftab{number_coefficients}. Assume that the elements in $\vec{\beta}$ (corresponding to the rows in $\vec{B}_\ifaceapprox$) are sorted in ascending order with respect to the corresponding powers of $\vec{t}$. Looping over all $K$ rows in $\vec{B}_\ifaceapprox$, the $m$-th row is discarded if it is linear dependent on the $m-1$ previous rows. The polynomials whose coefficients cannot be eliminated are of higher order, i.e.\ $\mathcal{O}(\norm{\vec{t}}^{2N+2})$. In the limiting case $\curv_1=\curv_2$, this approach produces the analytical solution given in \refeqn{surfaceLB_approx_solution_3D}. We would like to emphasze that due to $\operatorname{rank}(\vec{B}_\ifaceapprox)=(N+1)(M+1)$, the reduced form of \refeqn{polynomial_comparison_linear_system} can be solved exactly. \ref{app:comparison_explicit_notation} contains the full expansion of the first three entries of the coefficient vector.

\begin{table}[ht]
\centering
\caption{Number of coefficients $K$ over various $N=M$, with the apparent relation $K-(N+1)^2=4N+2$.}%
\label{tab:number_coefficients}%
\begin{tabular}{c|c|c|c|c|c|c}%
$N$&1&2&3&4&5&6\\%
\hline%
$K$&10&19&30&43&58&75\\%
\hline%
$K-(N+1)^2$&6&10&14&18&22&26 
\end{tabular}
\end{table}

\section{The numerical algorithm}\label{sec:numerical_algorithm}%
\refFig{flowchart_numerical_algorithm} contains a schematic flowchart\footnote{Note that due to to the cell based application parallelization of this algorithm is trivial.} of the developed numerical algorithm. The intersections $\vec{x}_{\partial\iface,k}$ of the true hypersurface $\iface$ with the cell edges are computed by \name{Newton} iteration. The level-set function is approximated by a third-order polynomial based on the values of the level-set function $\lvlset_\iface$ and its gradient $\grad{\lvlset_\iface}$, evaluated at the cell corners. If the hypersurface is parameterizable over some parameter set $\paradomain_\iface$, i.e.\ $\iface=\{\vec{g}_\iface(\vec{t}):\vec{t}\in\paradomain_\iface\}$, the centroid of the polygon spanned by the edge intersections is projected onto $\paradomain_\iface$ to obtain $\hat{\vec{t}}_{\partial\iface}:=\vec{g}^{-1}_\iface(\hat{\vec{x}}_{\partial\iface})$, with $\hat{\vec{x}}_{\partial\iface}=\nicefrac{1}{N}\sum_{k=1}^{N}{\vec{x}_{\partial\iface,k}}$ and $\vec{g}^{-1}_\iface:\setR^d\mapsto\paradomain_\iface$ surjective. The base point is then obtained as $\vec{x}_0=\vec{g}_\iface(\hat{\vec{t}}_{\partial\iface})$. For hypersurfaces that are not parameterizable in the above sense, a metric projection dependent on the class of the respective hypersurface is applied. The principal curvatures $\curv_i$ and associated directions $\vec{\tau}_i$ define the approximated hypersurface $\ifaceapprox$, whose intersections $\vec{x}_{\partial\ifaceapprox,k}$ with the cell edges, after projection onto the tangential plane via $\mathcal{P}_\ifaceapprox(\vec{x}):=[\vec{\tau}_1,\vec{\tau}_2]^{\sf T}(\vec{x}-\vec{x}_0)$, provide the vertices $\vec{t}_k$ of the parameter set polygon $\paradomain_{\ifaceapprox}\approx\bigcup_k{\triangle_k}$; cf. again \reffig{support_triangulation}. Due to the polynomial character of the underlying equation, cf.~subsection \ref{subsec:equating_polynomial_coefficients}, it is possible to approximate the solution either by a variational formulation or by comparison of polynomial coefficients. In the latter case, the coefficients associated to the ansatz functions $\ansatzfun[k]$ can be evaluated directly. The first case, i.e.\ the application of \name{Petrov-Galerkin} approach, however requires to assemble a linear system, which is solved employing the \name{LAPACK} routines \texttt{DGETRF} and \texttt{DGETRS}. Numerical experiments have shown that for very small hypersurface patches ($\lvert\ifaceapprox\rvert\leq\num{e-7}$)\footnote{This value is related to the tolerance employed in the root finding algorithm, where we used $\num{e-6}$.}, the system may become ill-posed. For those non-invertible matrices $\vec{A}_\ifaceapprox$, the principal curvatures $\curv_i$ are set to zero, corresponding to a planar approximation, and the edge intersections $\vec{x}_{\partial\ifaceapprox,k}$ are recomputed. After assembling the solution $u$, the rightmost expression of \refeqn{volume_integration_base_equation} can be evaluated. The integral over the hypersurface is evaluated using \refeqn{application_surface_divergence} on $\partial\ifaceapprox=\bigcup_k{\partial\ifaceapprox_k}$, where the details are given in subsection \ref{subsec:quadrature_curve_integral}. A cell face $\face_k$ with a non-zero contribution to \refeqn{volume_integration_base_equation} is either intersected by $\ifaceapprox$ or interior (i.e., $\lvlset_\ifaceapprox(\vec{x})<0\quad\forall\vec{x}\in\face_k$), where in the first case the area is computed by standard quadrature. If the computed volume is negative or exceeds the volume of the containing cell, the curvatures $\curv_i$ are set to zero, an the edge intersections $\vec{x}_{\partial\ifaceapprox,k}$ are recomputed as in the case of a non-invertible $\vec{A}_\ifaceapprox$. This case will be referred to as \textit{out of bounds} below.     

\begin{figure}[ht]
\centering%
\includegraphics{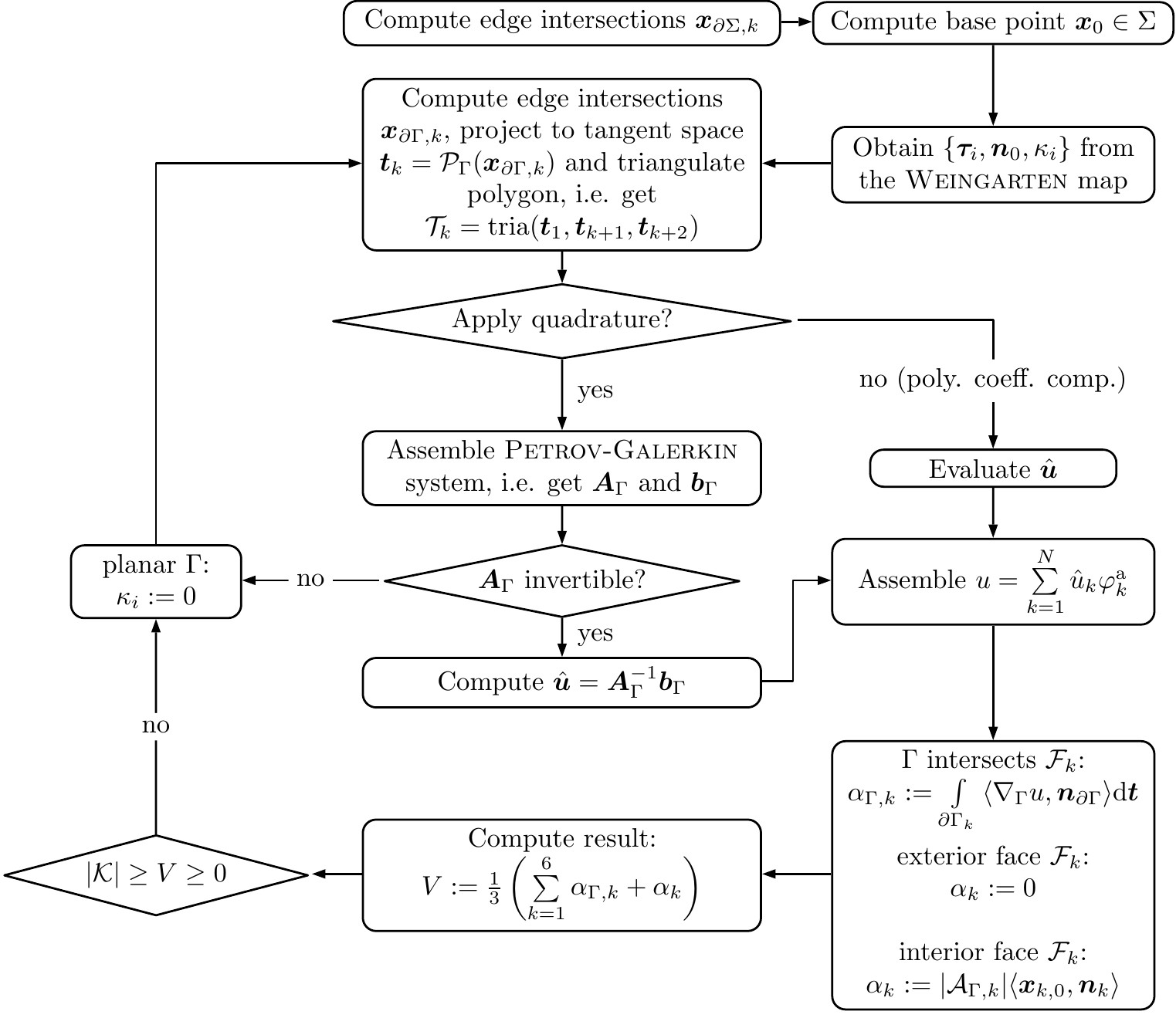}
\caption{Flowchart of the numerical algorithm.}%
\label{fig:flowchart_numerical_algorithm}%
\end{figure}
\subsection{Numerical quadrature of curve integrals}\label{subsec:quadrature_curve_integral}%
The present subsection is concerned with the evaluation of integrals of the form $\int_{\partial\ifaceapprox}{\iprod{\grads[\ifaceapprox]{u}}{\vec{n}_{\partial\ifaceapprox}}\dd{l}}$, where $u:\ifaceapprox\mapsto\setR$ is the numerical solution of \refeqn{surfaceLB_variational_formulation}. As stated above, cf.~\reffig{cell_intersection_status}, a boundary curve segment $\partial\ifaceapprox_k=\partial\ifaceapprox\cap\mathcal{F}_k$ contained in the rectangular face $\mathcal{F}_k$ can be parameterized in two ways. Firstly, in terms of a height function over one of the edges of the face $\mathcal{F}_k$, i.e.
\begin{align}
\partial\ifaceapprox_k=\{\vec{g}_{\partial\ifaceapprox,k}(\mu):\mu\in\paradomain_{\partial\ifaceapprox,k}\}\quad\text{with}\quad\vec{g}_{\partial\ifaceapprox,k}:=\vec{x}_{0,k}+\mu\vec{b}_{k}+h_{\partial\ifaceapprox,k}(\mu)\vec{n}_k,\label{eqn:boundary_segment_height_function}%
\end{align}
where $\paradomain_{\partial\ifaceapprox,k}$ is the simply connected parameter domain of the height function. Alternatively, polar coordinates can be applied, yielding
\begin{align}
\partial\ifaceapprox_k=\{\vec{g}_{\partial\ifaceapprox,k}(\mu):\mu\in[0,\nicefrac{\pi}{2}]\}\quad\text{with}\quad\vec{g}_{\partial\ifaceapprox,k}:=\vec{x}_{0,k}+r_{\partial\ifaceapprox,k}(\mu)\vec{e}_{r}(\mu),\label{eqn:boundary_segment_polar_function}%
\end{align}
where ${\vec{e}_r}_{\vert\mu=0}=\vec{b}_k$ and ${\vec{e}_r}_{\vert\mu=\pi}=\vec{n}_k$. The latter representation is chosen if two adjacent edges of a face are intersected, whereas the height function is used in the case of opposing intersected edges. The polar representation is required to cover the case where $\partial\ifaceapprox_k$ is not the graph of a function whose independent variable varies along an edge, cf.~\reffig{surfaceLB_evaluation_example_2D_01}. Since we ultimately wish to perform quadrature operations on $\partial\ifaceapprox_k$, the quadrature nodes $\mu_{k,i}$ need to be chosen carefully to ensure good approximation for strongly varying $r_{\partial\ifaceapprox,k}$. The standard \name{Gauss-Legendre} nodes $\mu_{i,k}\in[0,\nicefrac{\pi}{2}]$ are transformed via
\begin{align}
\tilde{\mu}_{k,i}=\tan^{-1}\left(\alpha_k\tan\mu_{k,i}\right),\label{eqn:boundary_node_transformation}%
\end{align}
where $\alpha_k$ denotes the ratio of the distances of the interface intersection $\vec{x}_{0,k}$ to the base point $\vec{x}_{\partial\ifaceapprox,k}$; cf.~\reffig{illustration_boundary_node_transformation} for an illustration.

\begin{figure}[ht]
\null\hfill%
\subfigure[without transformation]{\includegraphics[]{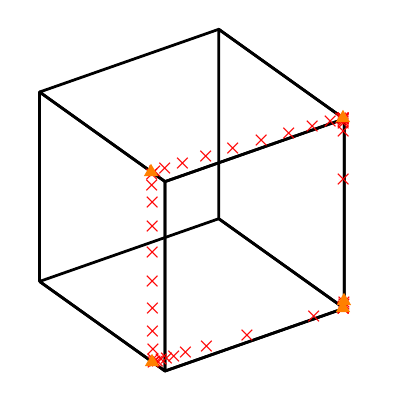}\label{fig:illustration_boundary_node_transformation_01}}%
\hfill
\subfigure[with transformation]{\includegraphics[]{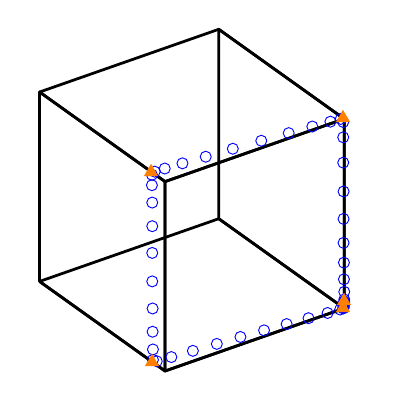}\label{fig:illustration_boundary_node_transformation_02}}%
\hfill
\subfigure[bottom face ($\mathcal{F}_3$) with quadrature nodes ($\alpha_3\approx\nicefrac{1}{19}$)]{\includegraphics[]{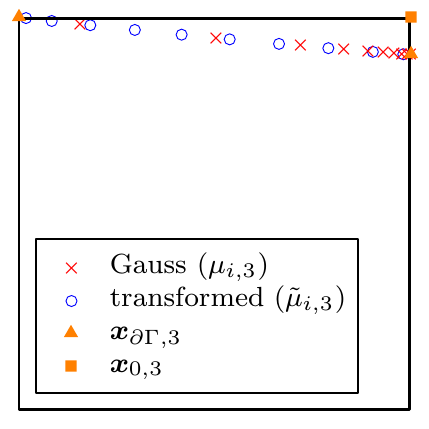}\label{fig:illustration_boundary_node_transformation_03}}%
\hfill\null%
\caption{Exemplified quadrature nodes on boundary curve $\partial\ifaceapprox$ with ($\color{blue}\circ$) and without ($\color{red}\times$) application of transformation; cf.~\refeqn{boundary_node_transformation}.}%
\label{fig:illustration_boundary_node_transformation}%
\end{figure}

The upcoming transformations are derived for the first case, their polar pendants can be obtained by analogous manner. First, note that the choice of the local coordinate system $\{\vec{b}_k,\vec{n}_k\}$ is not unique, in general. By equating the expressions in \refeqn{boundary_segment_height_function} and \refeqn{approx_hypersurface_graph_representation} and computing the appropriate inner products for $i\in\{1,2\}$, one obtains 
\begin{align}
t_i&=\iprod{\vec{x}_{0,k}-\vec{x}_0}{\vec{\tau}_i}+\mu\iprod{\vec{b}_k}{\vec{\tau}_i}+\iprod{\vec{n}_k}{\vec{\tau}_i}h_{\partial\ifaceapprox,k}=:\alpha_i+\mu\beta_i+\gamma_ih_{\partial\ifaceapprox,k},\label{eqn:boundary_height_implicit_01}\\%
\frac{1}{2}\left(\curv_1t_1^2+\curv_2t_2^2\right)&=\iprod{\vec{x}_{0,k}-\vec{x}_0}{\vec{n}_0}+\mu\iprod{\vec{b}_k}{\vec{n}_0}+\iprod{\vec{n}_k}{\vec{n}_0}h_{\partial\ifaceapprox,k}=:\alpha_3+\mu\beta_3+\gamma_3h_{\partial\ifaceapprox,k}.\label{eqn:boundary_height_implicit_02}%
\end{align}
Inserting \refeqn{boundary_height_implicit_01} in \refeqn{boundary_height_implicit_02} and rearranging yields the implicit quadratic relation
\begin{align}
c_{2,k}h_{\partial\ifaceapprox,k}^2+c_{1,k}(\mu)h_{\partial\ifaceapprox,k}+c_{0,k}(\mu)=0\label{eqn:boundary_segment_height_implicit}%
\end{align}
with coefficients
\begin{align}
\begin{split}
c_{0,k}&=\frac{\mu^2}{2}\left(\curv_1\beta_1^2+\curv_2\beta_2^2\right)+\mu\left(\curv_1\alpha_1\beta_1+\curv_2\alpha_2\beta_2-\beta_3\right)+\frac{1}{2}\left(\curv_1\alpha_1^2+\curv_2\alpha_2^2\right)-\alpha_3,\\%
c_{1,k}&=\mu\left(\curv_1\beta_1\gamma_1+\curv_2\beta_2\gamma_2\right)+\curv_1\alpha_1\gamma_1+\curv_2\alpha_2\gamma_2-\gamma_3,\\%
c_{2,k}&=\frac{1}{2}\left(\curv_1\gamma_1^2+\curv_2\gamma_2^2\right).%
\end{split}
\end{align}
Despite the possibility of explicitly calculating the roots of \refeqn{boundary_segment_height_implicit}, we prefer to apply a \name{Newton} algorithm. Also, the derivative of the height function $h_{\partial\ifaceapprox,k}$ with respect to $\mu$, which is required in \refeqn{curve_integral_transform} below for the integral transformation, can be computed by differentiating \refeqn{boundary_segment_height_implicit} and rearranging, i.e.
\begin{align}
\frac{\partial h_{\partial\ifaceapprox,k}}{\partial\mu}=-\frac{\partial_\mu c_{0,k}+h_{\partial\ifaceapprox,k}\partial_\mu c_{1,k}}{2h_{\partial\ifaceapprox,k}c_{2,k}+c_{1,k}}.%
\end{align}
The boundary normal emerges from the projection of the face normal $\vec{n}_{\mathcal{F},k}$ onto the tangent space, i.e.
\begin{align}
\vec{n}_{\partial\ifaceapprox,k}(\vec{t})=\frac{\vec{P}_\ifaceapprox\vec{n}_{\mathcal{F},k}}{\lVert\vec{P}_\ifaceapprox\vec{n}_{\mathcal{F},k}\rVert}.%
\end{align}
Finally, the curve integral is transformed as
\begin{align}
\int\limits_{\partial\ifaceapprox_k}{\iprod{\grads[\ifaceapprox]{u}}{\vec{n}_{\partial\ifaceapprox,k}}(\vec{t})\,\dd{l}}&=\int\limits_{\paradomain_{\partial\ifaceapprox,k}}{\iprod{\grads[\ifaceapprox]{u}}{\vec{n}_{\partial\ifaceapprox,k}}\left(\vec{g}_{\partial\ifaceapprox,k}(\mu)\right)\sqrt{1+\partial_\mu h_{\partial\ifaceapprox,k}^2}\,\dd{\mu}},\label{eqn:curve_integral_transform}
\end{align}
where the numerical evaluation is, once again, carried out by standard \name{Gauss-Legendre} quadrature. \refFig{quantities_quadrature_curve_integrals} illustrates the relevant quantities.
\begin{figure}[ht]
\null\hfill%
\includegraphics{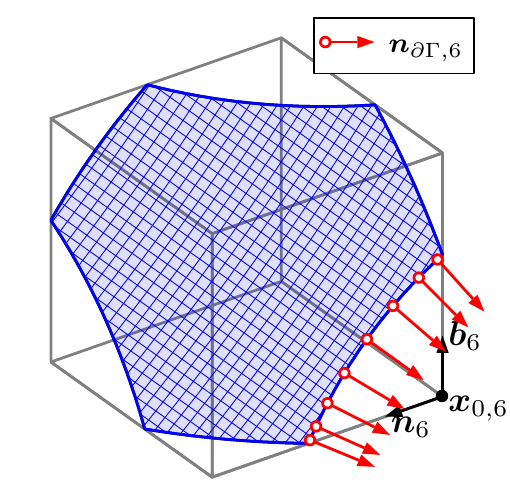}%
\hfill%
\includegraphics{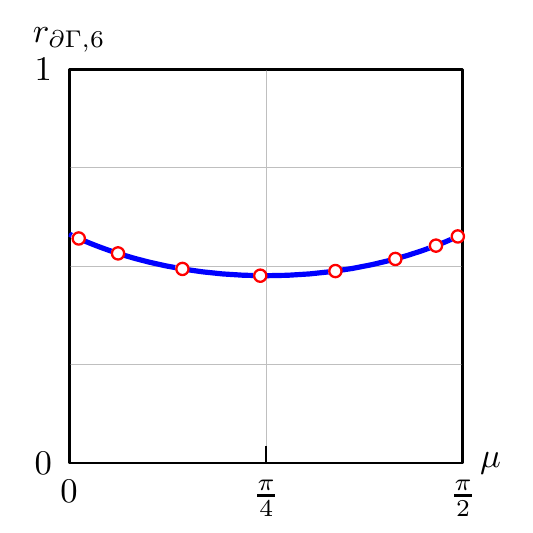}%
\hfill\null%
\caption{Relevant quantities for numerical quadrature of curve integrals, where the boundary normals are evaluated for $M=8$ quadrature nodes on the parameter set $\paradomain_{\partial\ifaceapprox,6}={[0,\frac{\pi}{2}]}$ and $\alpha_6=1$.}%
\label{fig:quantities_quadrature_curve_integrals}%
\end{figure}
\section{Numerical results}\label{sec:numerical_results}%
The present section gathers some numerical results for three classes of hypersurfaces, which are commonly encountered in the initial configuration of two-phase flow simulation: (i) ellipsoids with distinct and identical semi-axes $(a,b,c)$, the latter of course resembling spheres. (ii) hypersurfaces with rotational symmetry along the $z$-axis, whose radius is a quadratic function of the $z$-coordinate and (iii) perturbed spheres with base radius $R_0$ and variance $\sigma_0$.

Since the numerical evaluation of the original equation, cf.~\refeqn{surfaceLB_volume}, involves two significant distinguishable error sources, namely the approximation of the hypersurface and the numerical approximation of the variational problem, the convergence with increasing resolution is bounded by the approximation accuracy. Hence, due to the symmetry of local quadratic approximation of the hypersurface, one can obtain fourth-order convergence in space at most. The number of cells $\cell_i$ intersected by the hypersurface $\iface$ is denoted $N_\iface$, which is not an input parameter. In figures~\ref{fig:volume_error_ellipsoids} and \ref{fig:volume_error_liquid_bridges}, the computation of the referential error employs a discretization of the hypersurface parameter set $\paradomain_\iface$ into $N_\paradomain^2$ subdomains. In order to achieve comparability in terms of resolution, the errors produced by our algorithm are plotted over $\sqrt{N_{\iface}}$, approximately resembling the interface resolution per spatial dimension, i.e.\ $N_\paradomain^2\sim N_\iface$.
\subsection{Numerical setup}
The domain $\cell=[-1,1]^3$ under consideration is evenly discretized by $N_\cell$ cells per spatial dimension, where the center of volume coincides with the center of the domain. For both the \name{Petrov-Galerkin} (PG) and the polynomial comparison (PC) approach, the number of ansatz functions is $N\in\{4,9\}$, corresponding to products of even \name{Legendre} polynomials up to and including second ($M=2$) and fourth order ($M=4$), respectively; cf.~\refeqn{ansatz_function_definition}. The \name{Gauss-Legendre} quadrature of \refeqn{surfaceLB_variational_formulation} is carried out using $(2M+2)^2$ nodes per triangle, cf.~\refeqn{triangle_transformation_quadrature}. For the quadrature on the boundary curve segments $\partial\ifaceapprox_k$, cf.~\refeqn{curve_integral_transform}, $2M+6$ nodes are used. For all cases investigated below, there are no cells for which the coefficient matrix $\vec{A}_\ifaceapprox$ is non-invertible; cf.~\refeqn{polynomial_comparison_linear_system} and the flowchart in \reffig{flowchart_numerical_algorithm}. Moreover, for ellipsoids and hyperboloids of rotation no occurrences of \textit{out-of-bounds} are detected, while \reftab{perturbed_sphere_outofbounds} provides the numbers for perturbed spheres.  

\subsection{Referential volumes}
In order to separately analyze the error contribution of the local surface approximation, we compute the theoretical minimum error of the volume computation. The referential volumes are obtained as follows: the rectangular parameter set $\paradomain=[\alpha_1,\beta_1]\times[\alpha_2,\beta_2]$ is discretized in $N_\paradomain\times N_\paradomain$ equally sized rectangular pairwise disjoint subdomains $\paradomain_{ij}$, i.e.\ $\paradomain=\bigcup_{i,j=1}^{N_\iface}{\paradomain_{ij}}$ with
\begin{align}
\paradomain_{ij}=\left[\alpha_1+\frac{\beta_1-\alpha_1}{N_\iface}(i-1),\alpha_1+\frac{\beta_1-\alpha_1}{N_\iface}i\right]\times\left[\alpha_2+\frac{\beta_2-\alpha_2}{N_\iface}(j-1),\alpha_2+\frac{\beta_2-\alpha_2}{N_\iface}j\right].
\end{align}
The approximate interface patch $\ifaceapprox_{ij}$ is obtained by \name{Taylor} expansion of the height function $h_\iface$ around the respective center of $\paradomain_{ij}$, providing $\left\{\curv_i,\vec{\tau}_i,\vec{n}_0\right\}$. Next, we explicitly compute a set of $N_{\mathrm{quad}}^2=64$ quadrature weights and nodes $\{(\omega_k,\vec{x}_k)\}_{ij}$ with $\vec{x}_k\in\iface_{ij}$ which is projected to the approximate parameter set space, yielding $\{(\omega_k,\vec{b}_k)\}_{ij}$. Finally, the approximate volume $V^{\ifaceapprox}_{ij}$ is computed by evaluating the approximate height functions, while the true volume $V^{\iface}_{ij}$ is computed from the true height function, analytically where possible. For the hypersurfaces under consideration here, \reftab{volume_error_referential} gathers the relevant quantities. The global volume error then can be cast as
\begin{align}
\error_V^{\mathrm{ref}}:=\left\vert1-\left(\sum\limits_{i=1}^{N_\paradomain}{\sum\limits_{j=1}^{N_\paradomain}{V^{\ifaceapprox}_{ij}}}\right)\left(\sum\limits_{i=1}^{N_\paradomain}{\sum\limits_{j=1}^{N_\paradomain}{V^{\iface}_{ij}}}\right)^{-1}\right\vert\label{eqn:volume_error_referential},%
\end{align}
i.e.\ \refeqn{volume_error_referential} resembles an "upper bound" for the accuracy of the numerical implementation. 
\begin{table}[ht]%
\renewcommand{\arraystretch}{1.5}%
\centering%
\caption{Analytical volume segments for hypersurfaces under consideration for the numerical experiments, where the evaluation is carried out analytically for ellipsoids and hyperboloids of rotation.}%
\label{tab:volume_error_referential}%
\begin{tabular}{c||c|c|c}
class&parameters&$\paradomain_\iface$&$V^{\iface}_{ij}$\\hline%
\textbf{hyperb. of rev.}&$(r_0,\Delta r)$&$[0,2\pi)\times[-1,1]$&$\int\limits_{z_i}^{z_{i+1}}{\int\limits_{\varphi_i}^{\varphi_{i+1}}{\frac{\left(r_o+\Delta rz^2\right)^2}{2}\dd{\varphi}}\dd{z}}$\\%
\hline%
\textbf{ellipsoid}&$(a,b,c)$&$[0,2\pi)\times[0,\pi]$&$\frac{abc}{3}(\cos\theta_{j}-\cos\theta_{j-1})(\varphi_{i+1}-\varphi_{i})$\\%
\hline
\textbf{pert. sphere}&$(R_0,\sigma_0)$&$[0,2\pi)\times[0,\pi]$&$\frac{1}{3}\int\limits_{\varphi_i}^{\varphi_{i+1}}{\int\limits_{\theta_j}^{\theta_{j+1}}{R^3\sin\theta\dd{\theta}}\dd{\varphi}}$\\%
\end{tabular}
\end{table}
\subsection{Ellipsoids}\label{subsec:results_ellipsoids}%
\refFig{volume_error_ellipsoids} gathers the global numerical volume error for ellipsoids with different semi-axes, where the black and orange full circles denote the referential error of \refeqn{volume_error_referential} and the error obtained by linear approximation of the hypersurface, respectively. In general, the relative error decreases with increasing spatial resolution, commencing from between $\num{e-3}$ and $\num{e-4}$ for the lowest resolution of $N_\cell=10$ and reaching $\num{e-8}$ for spheres and $\num{e-7}$ for true ellipsoids, i.e.\ those with different semi-axes, respectively. For the latter the experimental order of convergence varies between $\num{3.00}$ and $\num{4.36}$, where in the cases presented here larger variations of curvatures do not necessarily produce lower orders of convergence; cf.~\reffig{volume_error_ellipsoids_EOC_over_beta}.

For all cases considered here, the absolute error of the \name{Petrov-Galerkin} approach lies approx.~two orders of magnitude below the error induced by linear approximation, indicating the benefits of exploiting local curvature information. Also, the \name{Petrov-Galerkin} approach outperforms the polynomial comparison for $N=4$ ansatz functions. As \reffig{volume_error_ellipsoids} indicates, the polynomial comparison requires $N=9$ ansatz functions (PC9) to produce results equivalent to (PG4). Polynomial comparison with $N=4$ ansatz functions (PC4) in general exhibits second order convergence in space, with the absolute error being roughly one order of magnitude below the linear approximation. This is due to the non-local character of the weak formulation underlying the \name{Petrov-Galerkin} approach, allowing for partial compensation of the higher order terms, which are neglected within the polynomial comparison. Moreover, increasing the number of ansatz functions to $N=9$ (PG9) does not improve the accuracy of the \name{Petrov-Galerkin} approach, implying that terms of fourth order in $t_i$ do not contribute significantly to the solution $u(\vec{t};\curv_i)$, irrespective of the sign and value of the principal curvatures; in fact, the observations of this paragraph extend to all classes of hypersurfaces investigated in this section, see figures~\ref{fig:volume_error_liquid_bridges} and \ref{fig:volume_error_perturbed_sphere}.

For the spherical case, cf.~\reffig{volume_error_ellipsoids_04}, the \name{Petrov-Galerkin} approach produces the expected fourth order convergence. Also, there is virtually no difference between the \name{Petrov-Galerkin} approach and the polynomial comparison, both for $N=4$ and $N=9$ ansatz functions. This is to be expected because the ansatz space $\ansatzfunset[4]$ already contains the analytical solution for $\curv_1=\curv_2=\curv$, hence an expansion cannot increase accuracy, since $\ansatzfunset[9]\supset\ansatzfunset[4]$.

Let $\beta:=\nicefrac{c}{a}$ be the ratio of the smallest and largest semi-axis. \refFig{volume_error_ellipsoids_EOC_over_beta} depicts the experimental order of convergence as a function of $\beta\in[\nicefrac{1}{2},\nicefrac{95}{100}]$, comparing the uniform variation of one (\textit{oblate}) and two (\textit{prolate}) semi-axes. For both oblate and prolate ellipsoids, the experimental order of convergence is approx. four, virtually independent of $\beta$. For the oblate ellipsoid with $\beta=\nicefrac{3}{5}$, cf.~\reffig{volume_error_ellipsoids_01}, the experimental order of convergence for PG4 drops to 3 due to sporadic increments in the absolute error magnitude. However, since we obtain forth order convergence in space for both smaller and larger values of $\beta$, we can deduce that this is caused by disadvantageous cancellation of local errors. An advantageous pronouncement of the aforementioned effect occurs for $N_\cell=20$ in PC9, where obtain an absolute error of approx. \num{e-8}, as compared to approx. \num{e-6} for $N_\cell=30$.

\begin{figure}[ht]
\centering 
\includegraphics{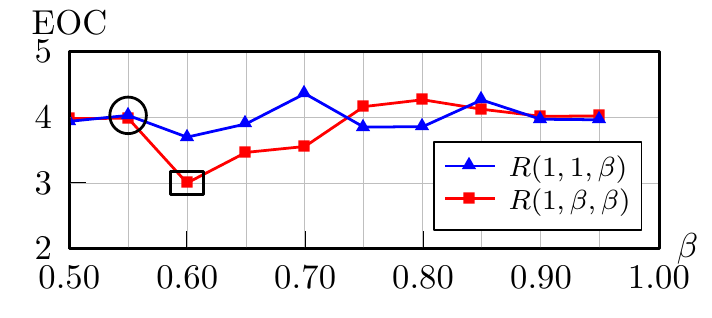}%
\caption{Experimental order of convergence of PG4 for $10\leq N_\cell\leq100$ cells per spatial direction for oblate ($\color{blue}\blacktriangle$) and prolate ($\color{red}\blacksquare$) ellipsoids over varying ratio of semi-axes ($R=0.99$). \refFig{volume_error_ellipsoids} below provides the underlying relative errors as a function of the spatial resolution $N_\cell$, where the points marked by the square/circle correspond to (\subref{fig:volume_error_ellipsoids_02}, \subref{fig:volume_error_ellipsoids_03}) / \subref{fig:volume_error_ellipsoids_01}.}%
\label{fig:volume_error_ellipsoids_EOC_over_beta}%
\end{figure}

\begin{figure}[h]
\null\hfill%
\subfigure[$\nicefrac{99}{100}(1,1,\nicefrac{3}{5})$]{
\includegraphics{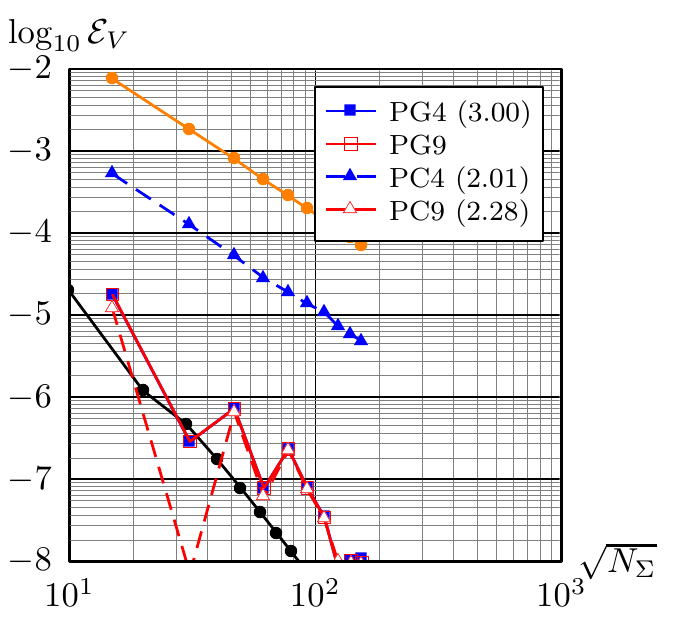}%
\label{fig:volume_error_ellipsoids_01}%
}%
\hfill%
\subfigure[$\nicefrac{99}{100}(1,\nicefrac{11}{20},\nicefrac{11}{20})$]{
\includegraphics{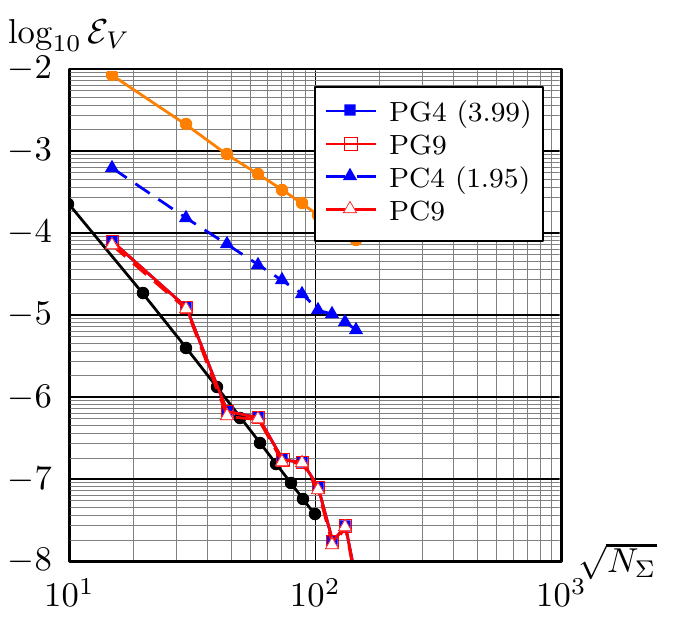}%
\label{fig:volume_error_ellipsoids_02}%
}%
\hfill\null%
\\
\null\hfill%
\subfigure[$\nicefrac{99}{100}(1,1,\nicefrac{11}{20})$]{
\includegraphics{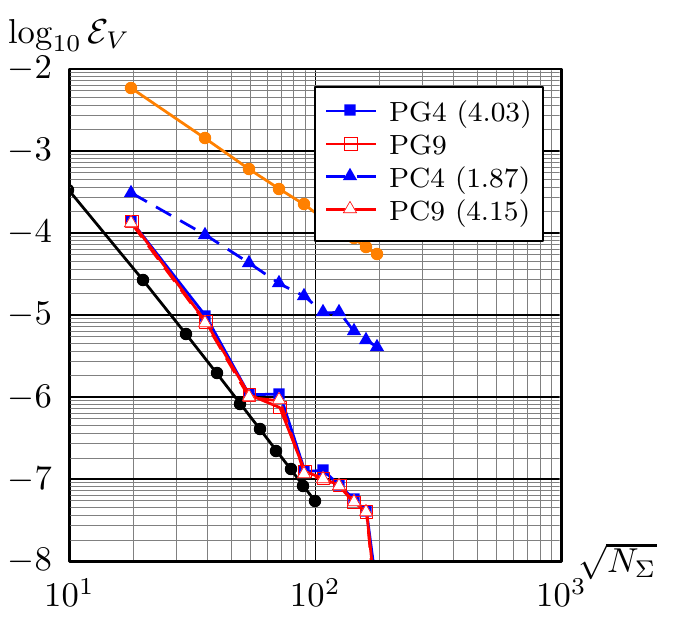}%
\label{fig:volume_error_ellipsoids_03}%
}%
\hfill%
\subfigure[$\nicefrac{99}{100}(1,1,1)$]{%
\includegraphics{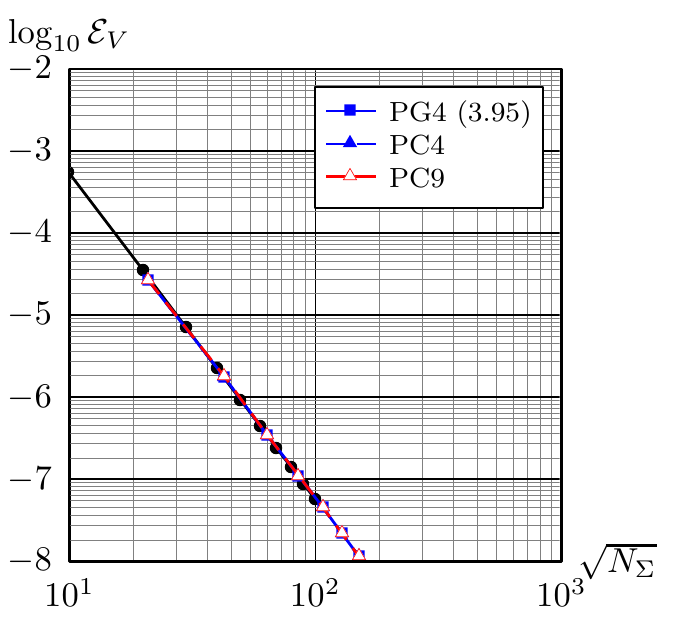}%
\label{fig:volume_error_ellipsoids_04}%
}%
\hfill\null%
\caption{Volume error (ellipsoids; subtitles denote semi-axes; number in brackets denote the EOC) over number of intersected cells $N_\iface$ with referential errors ($\bullet$: \refeqn{volume_error_referential} combined with \reftab{volume_error_referential}, $\color{orange}\bullet$: linear approximation of hypersurface). (PG) denotes the results obtained with the \name{Petriv-Galerkin} approach, (PC) refers to polynomial comparison.}%
\label{fig:volume_error_ellipsoids}%
\end{figure}
\subsection{Hyperboloids of revolution}\label{subsec:results_hyperboloids_revolution}%
Hyperboloids of revolution can be described by level-set functions of type
\begin{align}
\lvlset_\iface(\vec{x};r_0,\Delta r)=x^2+y^2-\left(r_0+\Delta r\,z^2\right)^2.%
\end{align}
\refFig{volume_error_liquid_bridges} shows the referential, cf.~\refeqn{volume_error_referential}, and numerical global volume error for hyperboloids of revolution with different radius variations. The observations concerning the evolution of the global error basically correspond to those of the ellipsoids. At this point, it is worth noting that we obtain fourth order convergence for (globally) non-convex hypersurfaces, cf.~figures~\ref{fig:volume_error_liquid_bridges_02} and \subref{fig:volume_error_liquid_bridges_04}.
\begin{figure}[h]
\null\hfill%
\subfigure[$(\nicefrac{\pi}{6},-\nicefrac{2}{5})$]{%
\includegraphics{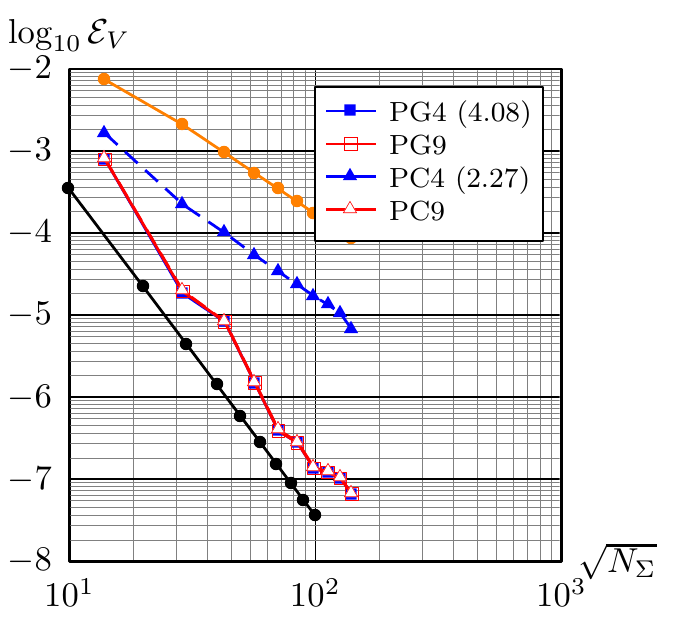}%
\label{fig:volume_error_liquid_bridges_01}}%
\hfill%
\subfigure[$(\nicefrac{\pi}{6},\nicefrac{2}{5})$]{%
\includegraphics{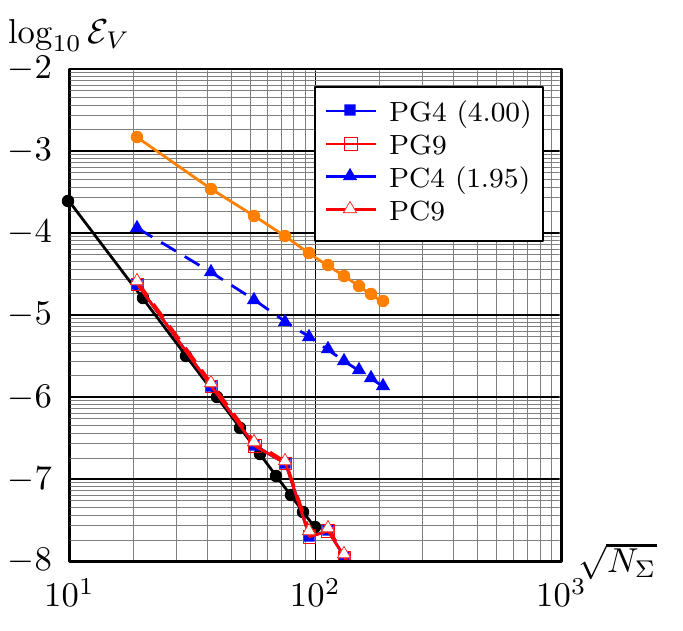}%
\label{fig:volume_error_liquid_bridges_02}}%
\hfill\null%
\\
\null\hfill%
\subfigure[$(\nicefrac{\pi}{6},-\nicefrac{2}{5})$]{%
\includegraphics[width=7cm]{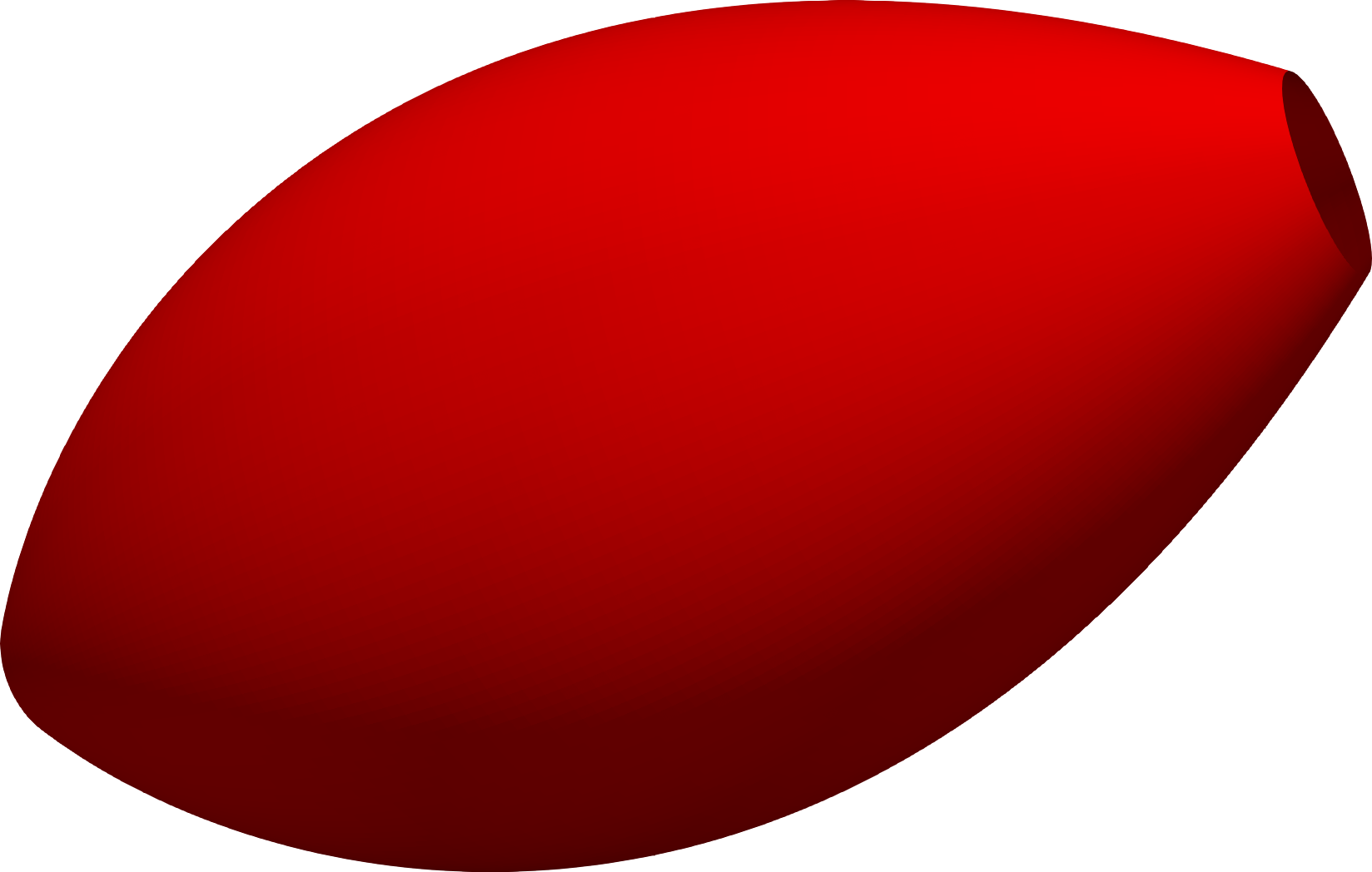}%
\label{fig:volume_error_liquid_bridges_03}}%
\hfill%
\subfigure[$(\nicefrac{\pi}{6},\nicefrac{2}{5})$]{%
\includegraphics[width=7cm]{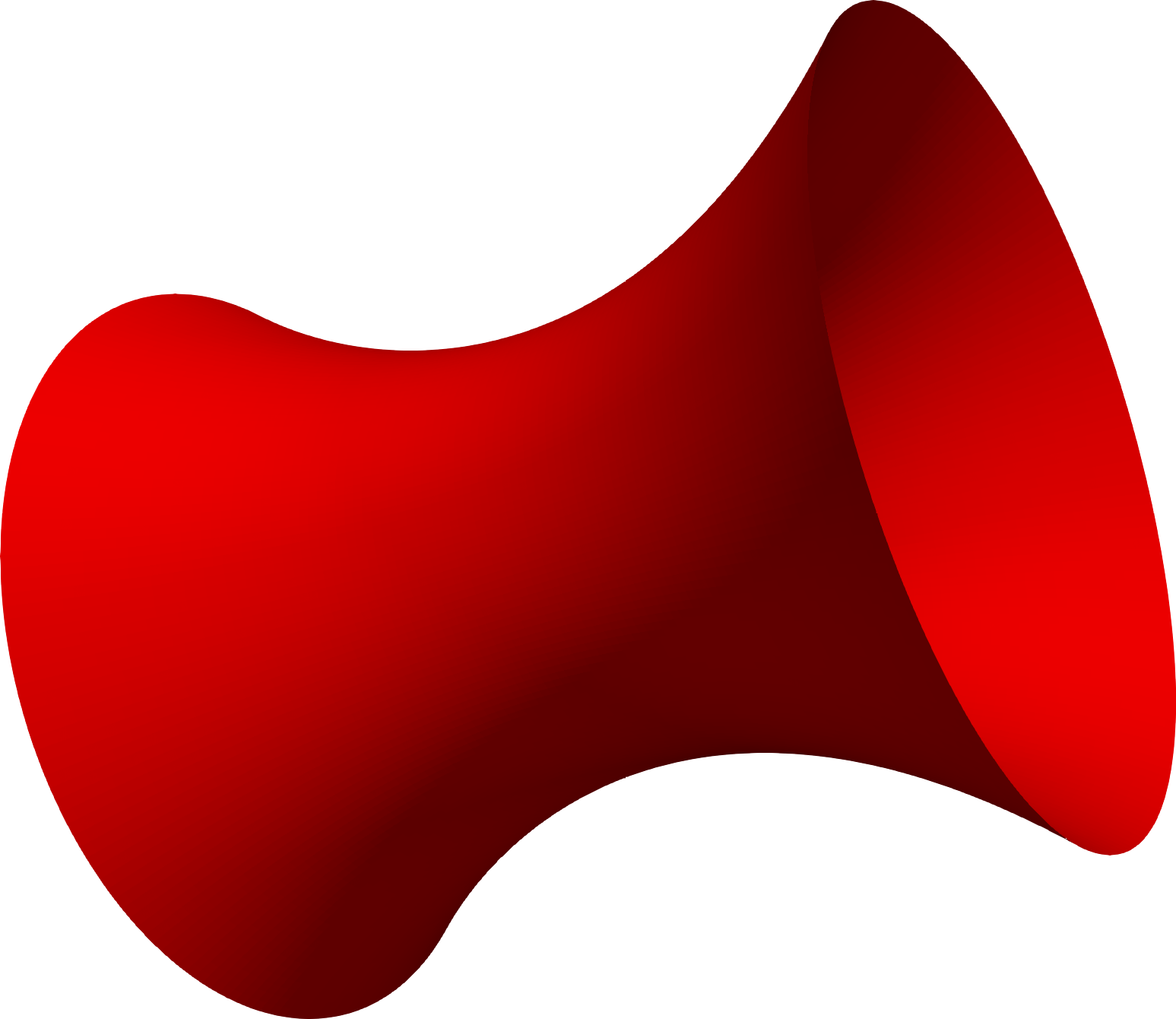}%
\label{fig:volume_error_liquid_bridges_04}}%
\hfill\null%
\caption{Volume error (hyperboloid of rotation; subtitles correspond to the parameters in \reftab{volume_error_referential}) over number of intersected cells $N_\iface$ with referential errors ($\bullet$: \refeqn{volume_error_referential} combined with \reftab{volume_error_referential}, $\color{orange}\bullet$: linear approximation of hypersurface) with illustrations.}%
\label{fig:volume_error_liquid_bridges}%
\end{figure}
\subsection{Perturbed spheres}\label{subsec:perturbed_spheres}%
Perturbed spheres can be described by level-set functions in spherical coordinates $\vec{r}:=[r,\varphi,\theta]^{\sf{T}}$ of type
\begin{align}
\lvlset_\iface(\vec{r};R_0,\sigma_0)=r^2-R^2(\varphi,\theta;R_0,\sigma_0),\label{eqn:perturbed_sphere_levelset}%
\end{align}
where the description of the radius $R$ employs tesseral spherical harmonics up to and including order $L\in\setN$, i.e.
\begin{align}
R^3=\sum\limits_{l=0}^{L}{\sum\limits_{m=-l}^{m=l}{c_l^m\mathcal{Y}_l^m(\varphi,\theta)}}.\label{eqn:perturbed_sphere_radius_series}%
\end{align}
The reason for expanding the third power of the radius instead of the radius itself is that the computation of the enclosed volume is considerably simplified, because $\lvert\operatorname{dom}(\iface)\rvert=c_0^0\nicefrac{\sqrt{4\pi}}{3}$. Moreover, to ensure continuity of the polar derivatives at the poles, modes with $m=\pm1$ are excluded, i.e.\ we enforce $c_l^{\pm1}\equiv0$; cf.~\ref{app:parametrization_hypersurface_spherical_harmonics} for details. The $(L+1)^2-2L$ coefficients $c_l^m\sim\mathcal{N}(0,\sigma_0)$ are computed by the \name{Box-Muller} method, i.e.
\begin{align}
c_l^m=%
\begin{cases}
\sqrt{4\pi}R_0^3&l=0\\%
\sqrt{\sigma_0}\sqrt{-2\log\gamma_1}\cos(2\pi\gamma_2)&l>0%
\end{cases}\qquad\text{with}\qquad\gamma_{1,2}\sim\mathcal{U}(0,1).%
\end{align}

In general, the observations concerning convergence and absolute error magnitude which have been established in \ref{subsec:results_ellipsoids} hold for the perturbed spheres as well. However, there are two characteristic differences. %
First, it is worth noting that the referential errors, cf.~\refeqn{volume_error_referential}, obtained from direct quadrature with $N_\mathrm{quad}=64$ nodes (see figures~\ref{fig:volume_error_perturbed_sphere_error_L03}, \subref{fig:volume_error_perturbed_sphere_error_L06} and \subref{fig:volume_error_perturbed_sphere_error_L09}) are larger than those obtained by application of our approach (excluding PC4), indicating its performance for locally non-convex hypersurfaces. If the deviation from the sphere is small, which is the case for $L=3$, the polynomial comparison performs better in terms of absolute error.
Second, while there were no cells whose volume fractions were out of bounds in \ref{subsec:results_ellipsoids} to \ref{subsec:results_hyperboloids_revolution}, this phenomenon occurs for perturbed spheres; cf.~\reftab{perturbed_sphere_outofbounds}. However, in the cases investigated here, the maximum number of those cells is three (obtained for $L=9$ with PC9), corresponding to \num{0.01}\% of the intersected cells; the affected cells share the property of having volume fractions close to 1 or 0\footnote{Note that the inverse relation is not true, i.e.\ cells with volume fractions close to 0 or 1 are generally not affected.}; cf.~\reffig{perturbed_sphere_oob_example} for details. This exceedance can be explained as follows: if all intersection points $\vec{x}_\iface$ are located in the very vicinity of corners, as illustrated in \reffig{perturbed_sphere_oob_example}, even small values of $d_\cell\curv_i$ can cause $\partial\ifaceapprox\not\in\cell$. In other words, evan small relative curvatures of the boundary curve potentially cause the latter to leave the cell under consideration. However, due to the aforementioned prerequisites concerning the intersection, this effect is expected to occur relatively rarely; cf.~again \reftab{perturbed_sphere_outofbounds}.      
\begin{figure}[h]
\null\hfill%
\includegraphics[width=5cm]{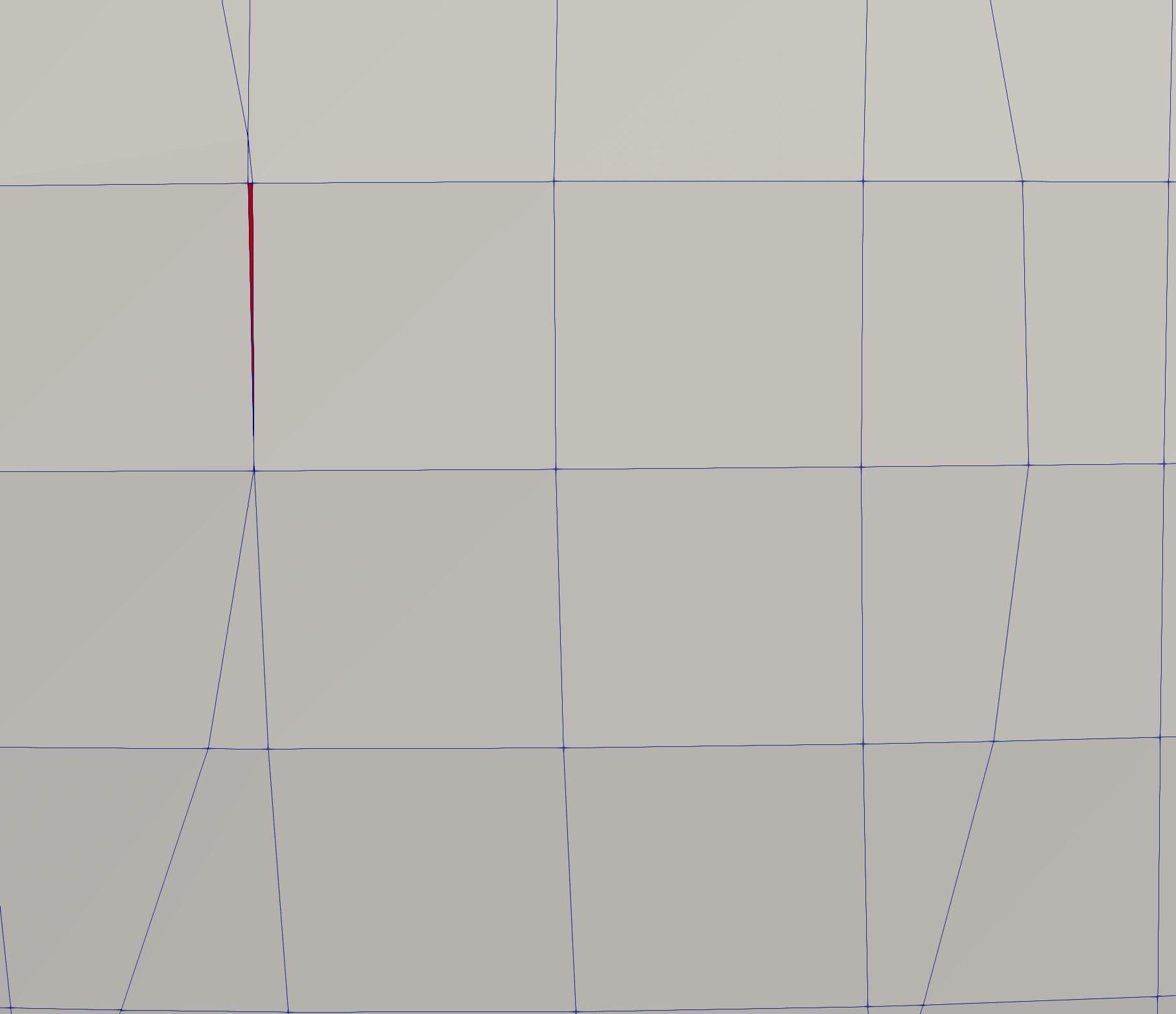}%
\hfill%
\includegraphics[width=3cm]{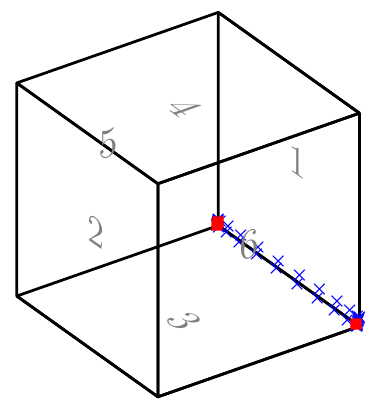}%
\hfill\null%
\caption{Left: surface mesh (red: out-of-bounds cells) for $L=9$ and $N_\cell=90$ (PG4 result: \num{-6.09e-11} and \num{-9.45e-8}); right: cell (volume \num{1.097e-5}) with hypersurface intersections $\vec{x}_{\iface}$  ($\color{red}\blacksquare$) and quadrature nodes $\vec{x}_{\partial\ifaceapprox}$ $(\color{blue}\times)$; cf.~\reftab{perturbed_sphere_outofbounds}.}%
\label{fig:perturbed_sphere_oob_example}%
\end{figure}

\begin{figure}[h]
\null\hfill%
\subfigure[rel. volume error]{%
\includegraphics[width=7cm]{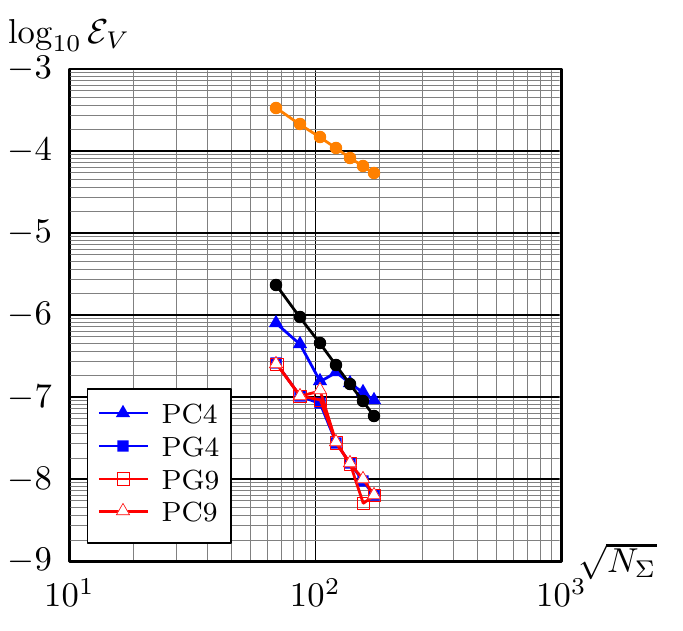}%
\label{fig:volume_error_perturbed_sphere_error_L03}}%
\hfill%
\subfigure[$\curv_1+\curv_2=2\curv_\iface\in{[-2.7,-2.2]}$]{%
\includegraphics[width=5cm]{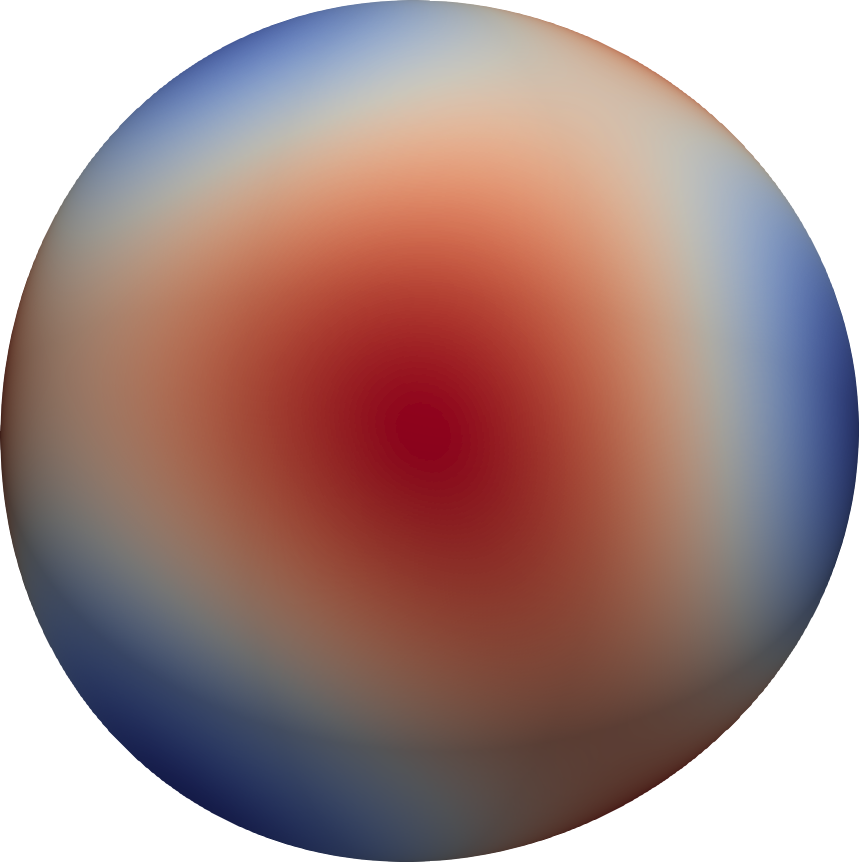}%
\label{fig:volume_error_perturbed_sphere_curvature_L03}}%
\hfill\null%
\\
\null\hfill%
\subfigure[rel. volume error]{%
\includegraphics[width=7cm]{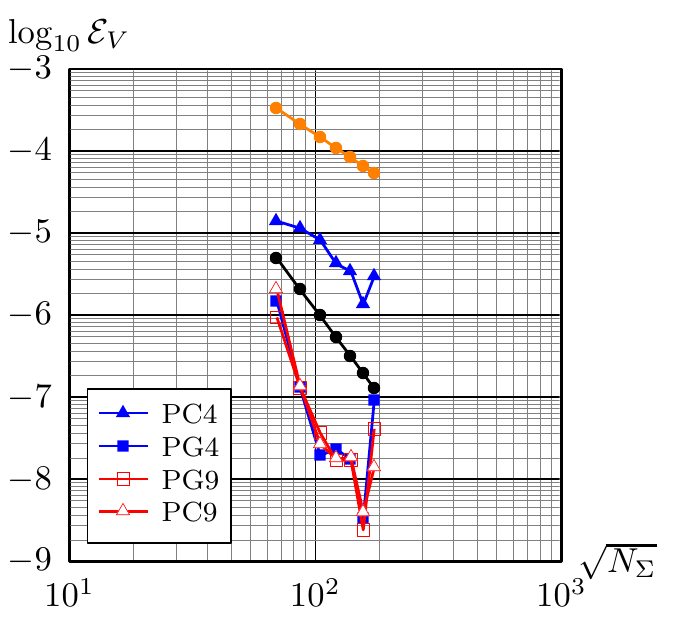}%
\label{fig:volume_error_perturbed_sphere_error_L06}}%
\hfill%
\subfigure[$\curv_1+\curv_2=2\curv_\iface\in{[-4.3,0.25]}$]{%
\includegraphics[width=5cm]{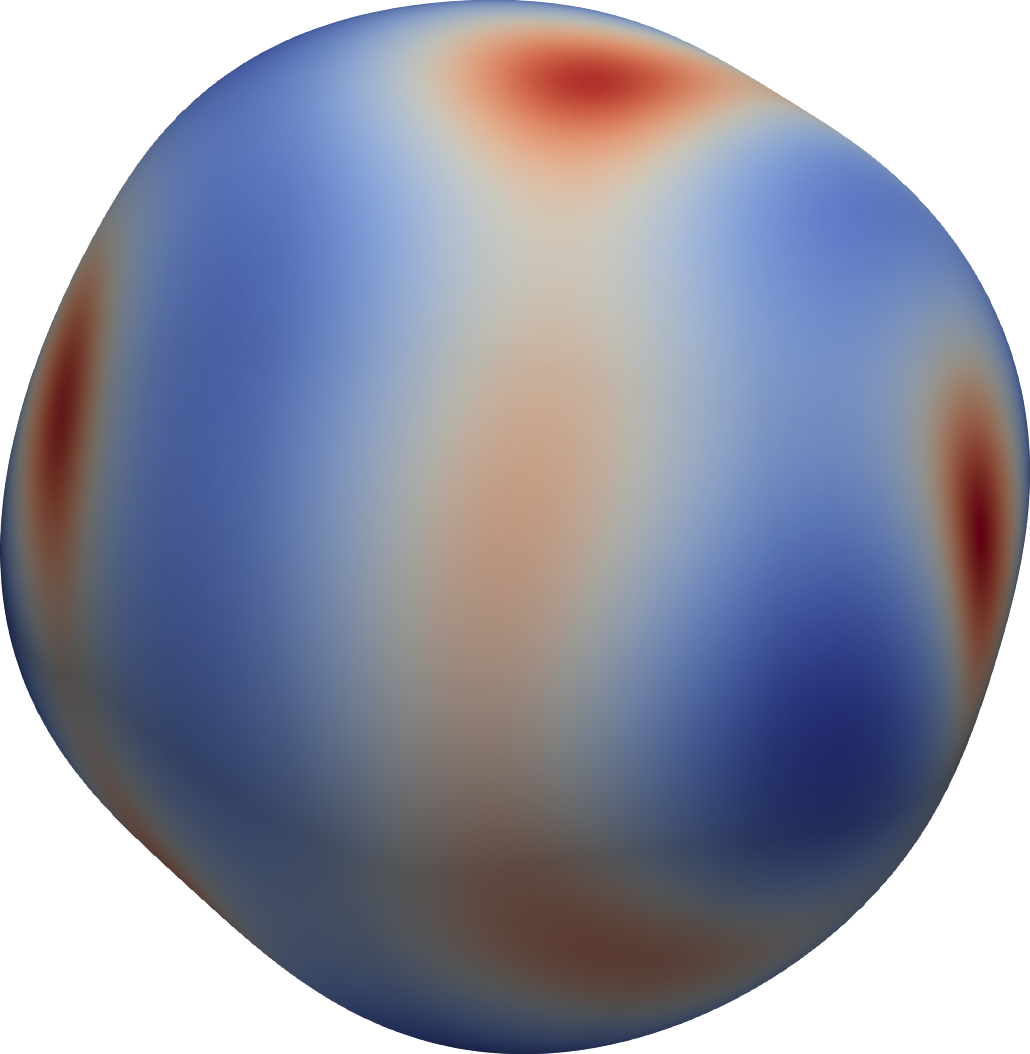}%
\label{fig:volume_error_perturbed_sphere_curvature_L06}}%
\hfill\null%
\\
\null\hfill%
\subfigure[rel. volume error]{%
\includegraphics[width=7cm]{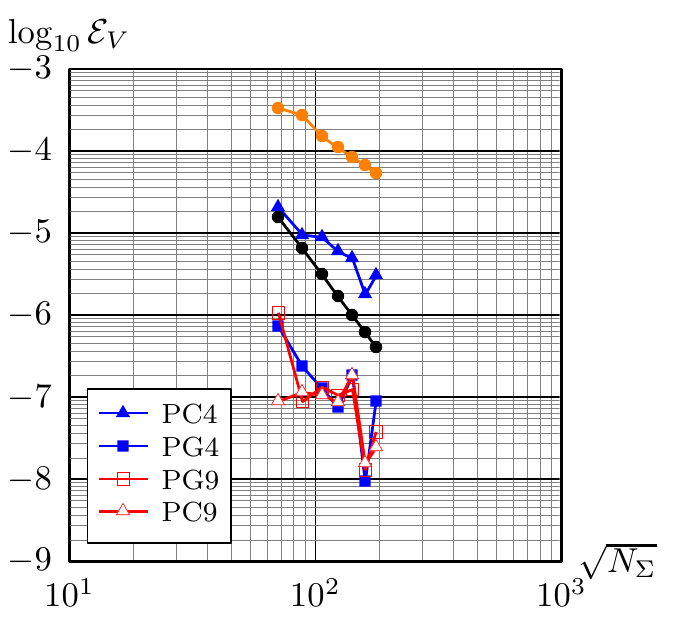}%
\label{fig:volume_error_perturbed_sphere_error_L09}}%
\hfill%
\subfigure[$\curv_1+\curv_2=2\curv_\iface\in{[-7,12]}$]{%
\includegraphics[width=5cm]{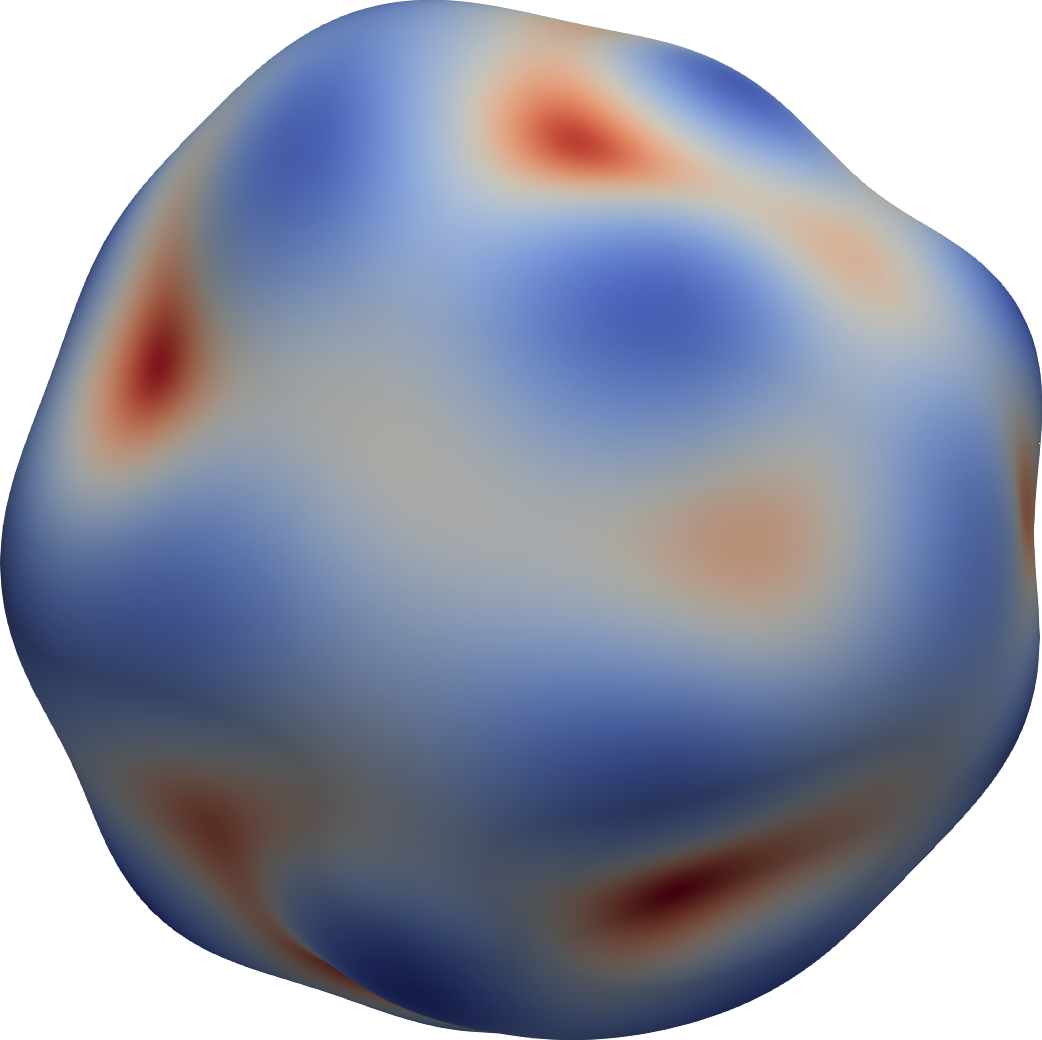}%
\label{fig:volume_error_perturbed_sphere_curvature_L09}}%
\hfill\null%
\caption{Volume error (perturbed sphere with $R_0=0.8$ and $\sigma_0=\num{5e-4}$; top to bottom row: $L\in\{3,6,9\}$) over number of intersected cells $N_\iface$ with referential errors ($\bullet$: \refeqn{volume_error_referential} combined with \reftab{volume_error_referential}, $\color{orange}\bullet$: linear approximation of hypersurface) with illustrations. The blue (red) regions in \subref{fig:volume_error_perturbed_sphere_curvature_L09} correspond to negative (positive) mean curvature $2\curv_\iface=\curv_1+\curv_2$.}%
\label{fig:volume_error_perturbed_sphere}%
\end{figure}

\begin{figure}[h]
\null\hfill%
\subfigure[$L=6$]{%
\includegraphics[width=5cm]{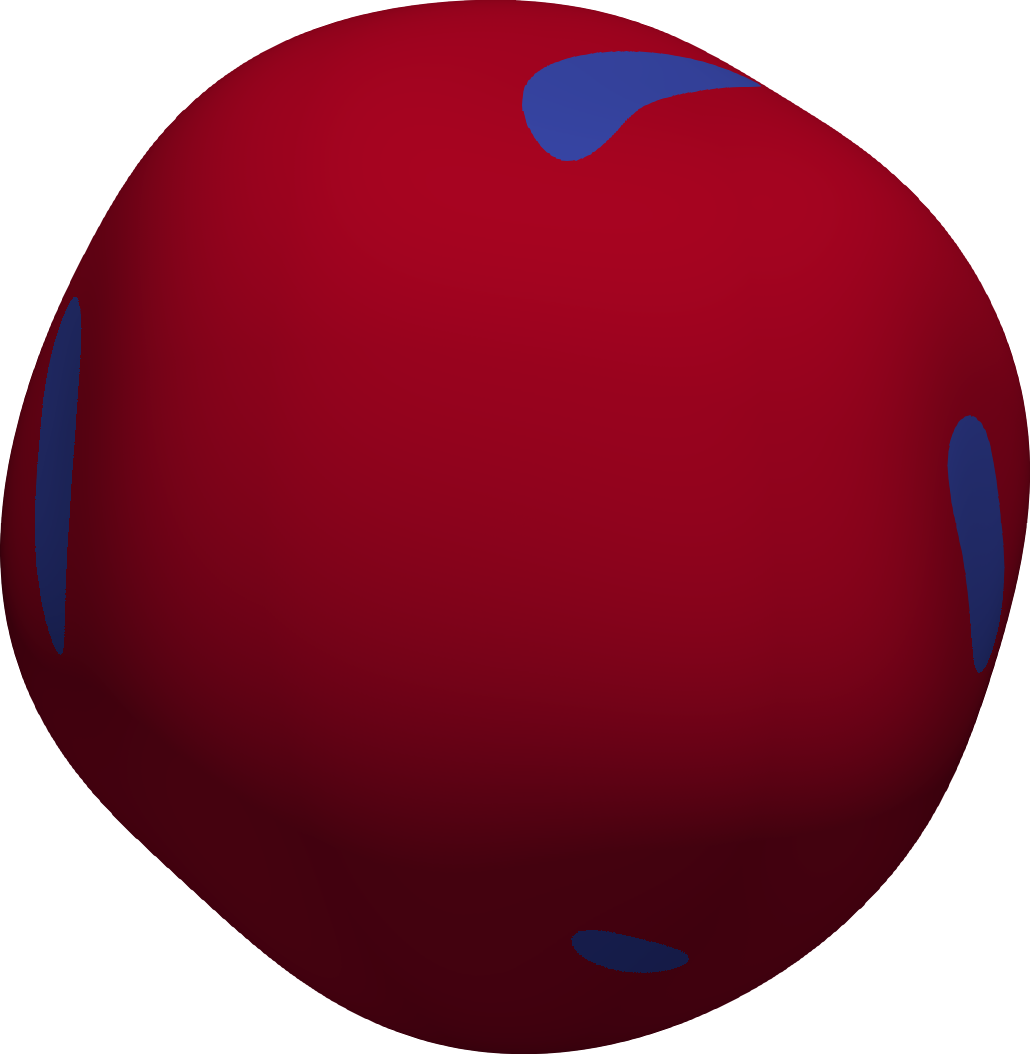}%
\label{fig:volume_error_perturbed_sphere_convexity_L06}}%
\hfill%
\subfigure[$L=9$]{%
\includegraphics[width=5cm]{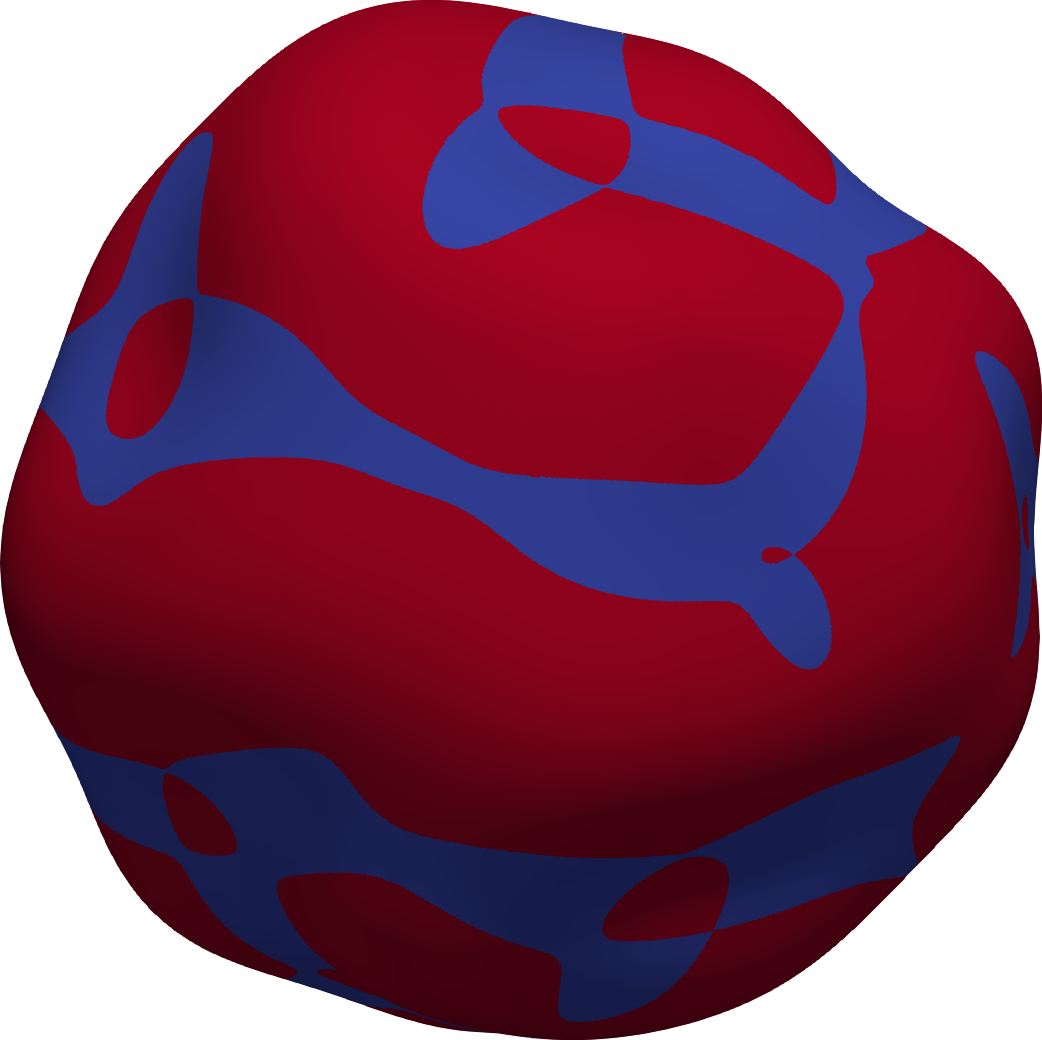}%
\label{fig:volume_error_perturbed_sphere_convexity_L09}}%
\hfill\null%
\caption{(\subref{fig:volume_error_perturbed_sphere_convexity_L06},\subref{fig:volume_error_perturbed_sphere_convexity_L09}): convexity index $\operatorname{sign}(\curv_1\curv_2)$ (red: \num{1}, blue: \num{-1}) for perturbed spheres with $L\in\{6,9\}$ evaluated on an evenly spaced discretization ($2000\times1000$) of the parameter domain $\mathbb{S}$.}%
\label{fig:perturbed_sphere_convexity}%
\end{figure}

\begin{table}[h]
\centering%
\caption{Number of interface cells of perturbed sphere for which $f_{k}\not\in{[0,1]}$ (\textit{out-of-bounds}; cf.~the flowchart in \reffig{flowchart_numerical_algorithm}) and total number of intersected cells $N_\iface$ over number of cells per spatial dimension $N_\cell$. Note that the lowest possible resolution is $N_\cell=40$.}%
\label{tab:perturbed_sphere_outofbounds}%
\input{./figures/tab/outofbounds.tex}\end{table}%
\section{Conclusion}\label{sec:conclusion}%
We have introduced an algorithm capable of computing volumes of domains which emerge from the intersection of cuboids and implicitly given hypersurfaces, where the novelty of the approach consists in the explicit exploitation of curvature information, i.e.\ principal curvatures and axes, in combination with the application of surface divergence theorem, where the solution of the emerging PDE is approximated by means of a \name{Petrov-Glaerkin} ansatz. The following main conclusions are drawn:
\begin{enumerate}
\item The local approximation of second order, exploiting geometrical (i.e.\ principal curvature) information from the \name{Weingarten} map, allows to obtain fourth-order convergence with spatial resolution. For all cases considered here, the absolute error is approximately three orders of magnitude below the error obtained by linear approximation of the hypersurface.
\item Fourth-order convergence is obtained for both convex and (globally and locally) non-convex hypersurfaces.
\item The proposed \name{Petrov-Galerkin} approach outperforms the polynomial comparison for an equal number of ansatz functions in terms of the absolute error, on average by one order of magnitude. Moreover, the results are robust with respect to the size of hypersurface patches, corresponding to the size of the parameter domains of the quadrature; cf.~\reffig{perturbed_sphere_surfacemesh} for an illustration.  
\item If (i) the principal curvatures are identical or (ii) one of the principal curvatures is zero, there is an analytical solution to the \name{Laplace-Beltrami} equation, which allows to compute the volume integrals exactly (with respect to the approximated hypersurface). This also considerably reduces the computational effort.
\end{enumerate}

\begin{figure}[h]
\null\hfill%
\subfigure[surface mesh]{%
\includegraphics[width=5cm]{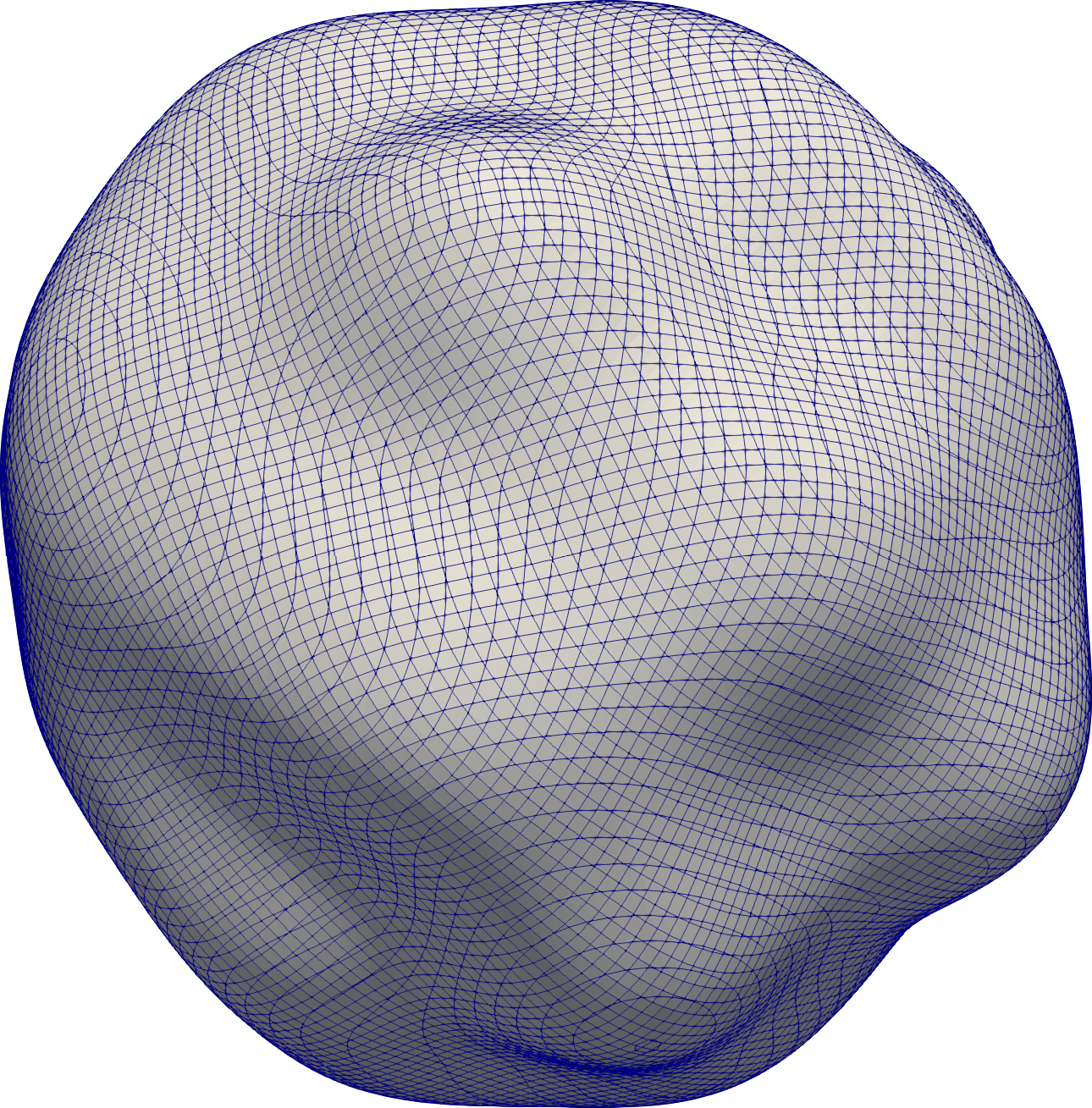}%
\label{fig:volume_error_perturbed_sphere_surfacemesh_global_L09}}%
\hfill%
\subfigure[close up]{%
\includegraphics[width=5cm]{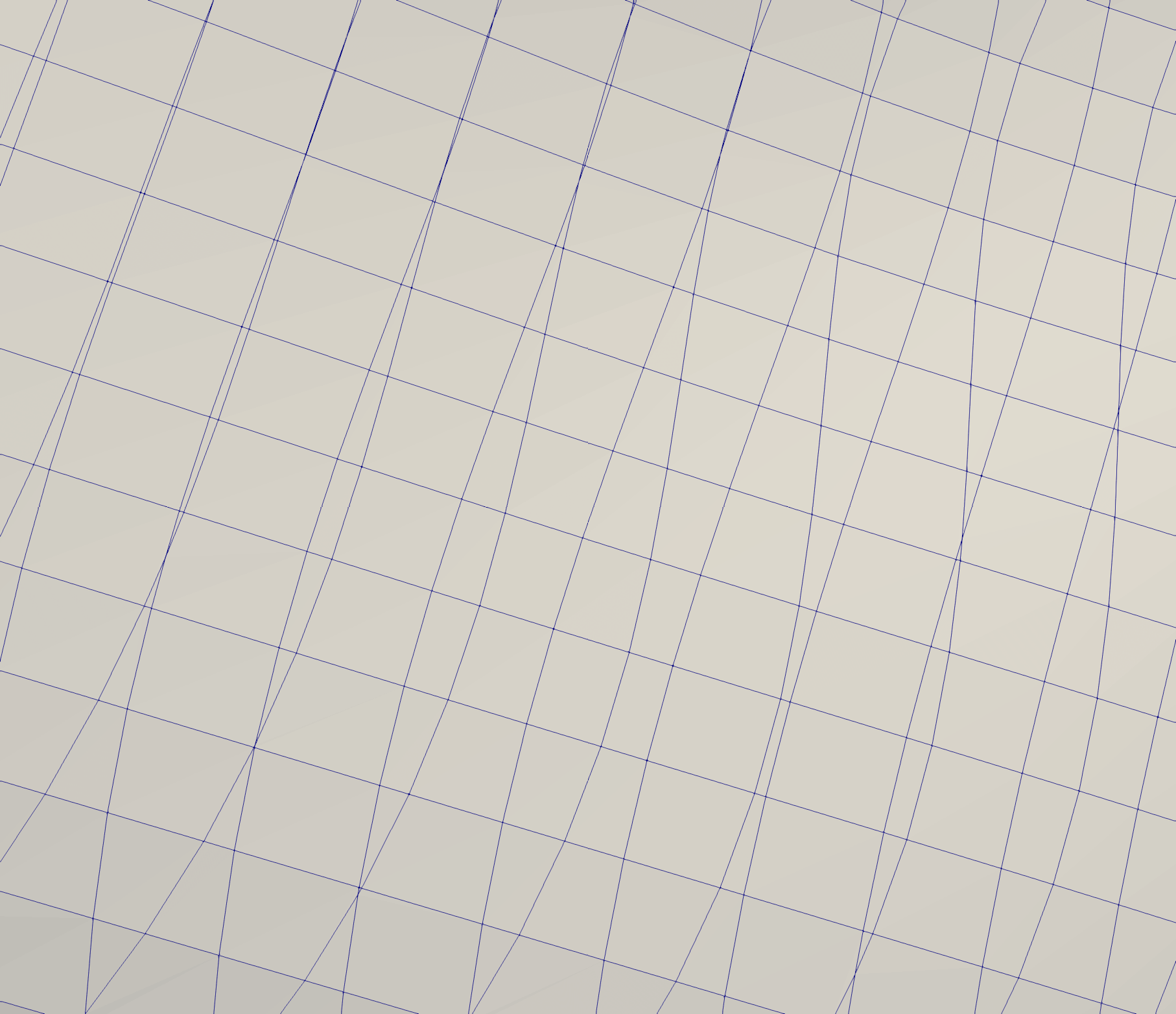}%
\label{fig:volume_error_perturbed_sphere_surfacemesh_local_L09}}%
\hfill\null%
\caption{Surface mesh for $L=9$ where the cuboidal domain was discretized in $N_\cell=100$ cells per spatial direction. The close up displays hypersurface patches with a wide range of sizes.}%
\label{fig:perturbed_sphere_surfacemesh}%
\end{figure}

In an upcoming paper, the authors will extend the numerical methods presented above for other right-hand sides of \refeqn{surfaceLB_volume}, especially for polynomials and constants, including the computation of surface area as an important special case.
\clearpage
\renewcommand{\bibname}{References}
\bibliographystyle{plainnat}
\bibliography{lbquad}
\nocite{*}
\begin{center}
\name{Acknowledgement}\\[2ex]%
The work of J. Kromer has been partly supported by the \textbf{Excellence Initiative of the German Federal and State Governments} and the \textbf{Graduate School of Computational Engineering} at Technical University Darmstadt, Germany. Also, the authors gratefully acknowledge financial support provided by the German Research Foundation (DFG) within the scope of \href{www.sfbtrr75.de}{SFB-TRR 75}.
\end{center}
\clearpage
\begin{appendix}
%
%
\section{A brief review of basic facts from  differential geometry}\label{app:review_differential_geometry}%
\citet{DCDS_2013_otmo} give a survey on smooth closed hypersurfaces embedded in $\setR^N$, including rigorous mathematical statements on the associated operators, fundamental forms and other geometrical properties. The present subsection heavily draws from their work. Here, however, we only reproduce those results needed within the scope of this work. For further mathematical details, the interested reader is referred to, e.g., the book of \citet{DiffGeo2005}.

Let $\iface$ be a hypersurface patch of class $\mathcal{C}^2$ confined by $\cell\subset\setR^d$ with $d\in\{2,3\}$, which is given via the zero iso-contour of a level-set $\phi$, i.e.
\begin{align}
\iface=\{\vec{x}\in\cell:\phi(\vec{x})=0\}\qquad\text{with normal}\qquad\ns:=\frac{\grad{\phi}}{\lVert\grad{\phi}\rVert}.\label{eqn:hypersurface_definition_levelset}%
\end{align}
Recall that we have $\partial\cell\supset\partial\iface\neq\emptyset$, by assumption; cf.~\reffig{cell_intersection_status}.%
\paragraph{Curvatures, first \& second fundamental form}%
For any point $\vec{x}_0\in\iface$ with outer unit normal $\ns$, there is a ball $\mathcal{B}_R(\vec{x}_0)\subset\setR^d$ with radius $R$ and a diffeomorphism $\Phi:\mathcal{B}_R(\vec{x}_0)\mapsto\mathcal{U}\subset\setR^d$, such that $\mathcal{U}\ni\Phi(\vec{x}_0)=\vec{0}$ and
\begin{align}
\Phi^{-1}(\mathcal{U}\cap(\setR^{d-1}\times\{0\}))=\Phi^{-1}(\paradomain_\iface\times\{0\}))=\mathcal{B}_R(\vec{x}_0)\cap\iface.\label{eqn:implicit_function_theorem}%
\end{align}
The implication of \refeqn{implicit_function_theorem} is that in the vicinity of $\vec{x}_0$, i.e.\ for $\lVert\vec{x}-\vec{x}_0\rVert\leq R$, the hypersurface can be parametrized over some (open) parameter set $\paradomain_\iface\subset\setR^{d-1}$, i.e.
\begin{align}
\iface\cap\mathcal{B}_R(\vec{x}_0)=\vec{g}(\paradomain_\iface;\vec{x}_0)\qquad\text{with}\qquad\vec{g}(\vec{t};\vec{x}_0):=\Phi^{-1}(\vec{t},0)\quad\text{and}\quad\vec{t}=\sum\limits_{i=1}^{d-1}{t_i\vec{e}_i};\label{eqn:surface_parametrization}%
\end{align}
cf.~\reffig{implicit_function_theorem} for an illustration. 
\begin{figure}[h]
\centering%
\includegraphics{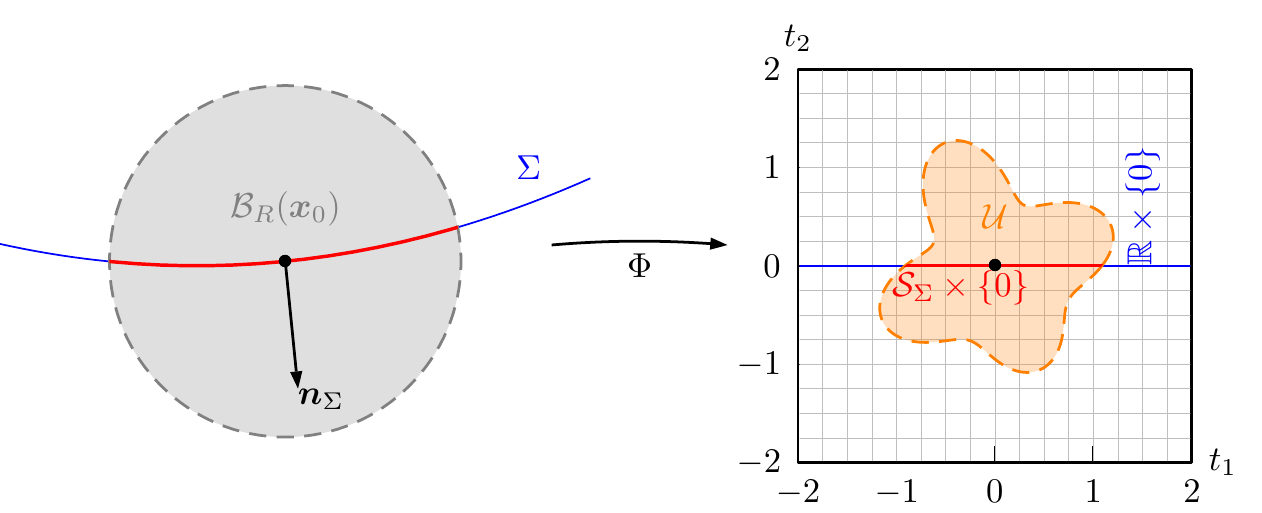}
\caption{Illustration of the implicit function theorem for $\iface\subset\setR^2$.}%
\label{fig:implicit_function_theorem}%
\end{figure}
The (covariant) tangent space $T_\iface$ attached to $\vec{x}_0$ is spanned by
\begin{align}
\vec{\nu}_i:=\frac{\partial\vec{g}}{\partial t_i}(\vec{t};\vec{x}_0)\qquad\text{for}\qquad 1\leq i\leq d-1,%
\end{align}
where $\iprod{\vec{\nu}_i}{\ns}\equiv0$, but, in general, $\lVert\vec{\nu}_i\rVert\neq1$ and $\iprod{\vec{\nu}_i}{\vec{\nu}_j}\neq0$. Analogously, $\vec{\nu}_{ij}$ denotes the second derivatives. Employing the \name{Einstein} summation convention, the \textit{first} and \textit{second fundamental form}, respectively, can be written as
\begin{align}
\vec{G}(\vec{x}_0):=\iprod{\vec{\nu}_i}{\vec{\nu}_j}\,\vec{e}_i\otimes\vec{e}_j%
\qquad\text{and}\qquad%
\vec{L}(\vec{x}_0):=\iprod{\ns}{\vec{\nu}_{ij}}\,\vec{e}_i\otimes\vec{e}_j.%
\end{align}
The eigenvalues $\{\curv_i\}\subset\setR^{d-1}$ of the \name{Weingarten} map $\vec{W}(\vec{x}_0):=\vec{G}^{-1}\vec{L} $, also called \textit{shape matrix} of $\iface$, correspond to the principal curvatures of the hypersurface  at $\vec{x}_0$. The associated eigenvectors $\vec{\tau}^0_i\in\setR^{d-1}$ provide the local directions of principal curvature, whose global pendant is obtained via $\vec{\tau}_i:=\iprod{\vec{\tau}^0_i}{\vec{e}_k}\vec{\nu}_k$ and normalization. Note that $\iprod{\vec{\tau}_i}{\vec{\tau}_j}=\delta_{ij}$ as well as $\iprod{\vec{\tau}_i}{\ns}=0$, i.e.\ $\{\vec{\tau}_i,\ns\}(\vec{x}_0)$ forms an orthonormal system and $T_\iface(\vec{x}_0)=\operatorname{span}(\vec{\tau}_i)$.
\paragraph{Surface gradient \& surface divergence}%
Let $f:\iface\mapsto\setR$ be a continuously differentiable field. Assume for the moment that the full gradient $\grad{f}$ exists. Then, the surface gradient can be understood as the projection of $\grad{f}$ onto the tangent space $T_\iface(\vec{x}_0)$, i.e.
\begin{align}
\grads{f}=\left(\ident-\ns\otimes\ns\right)\grad{f}=\frac{\dd{f}}{\dd{\vec{\tau}_2}}\vec{\tau}_1+\frac{\dd{f}}{\dd{\vec{\tau}_2}}\vec{\tau}_2.%
\end{align}
Note that left multiplication with $\PS:=\ident-\ns\otimes\ns$ corresponds to a projection onto the tangent plane $T_\iface(\vec{x}_0)$. Following \citet{ellPDE2001}, the lack of definition of the normal component can be eliminated by an extension of the definition, i.e.\ $f(\vec{x}\pm\epsilon\ns)=f(\vec{x})$ for $\vec{x}\in\iface$ and $\setR\ni\epsilon\ll1$. Hence, within a tubular neighborhood of thickness $2\epsilon$, the function value is extended to be constant along a normal deviation from the hypersurface. In the remainder of this paper, we assume any function mapping from the hypersurface $\iface$ to be extensible in this way. Then, the derivative in normal direction indeed becomes zero, since
\begin{align}
\frac{\dd{f}}{\dd{\ns}}=\lim\limits_{\epsilon\to0}\frac{f(\vec{x}+\epsilon\ns)-f(\vec{x})}{\epsilon}\equiv0.%
\end{align}
By analogous arguments, one obtains the surface divergence of a vector field $\vec{f}:\iface\mapsto\setR^d$ as
\begin{align}
\divs{\vec{f}}=\operatorname{tr}\left(\PS\grad{\vec{f}}\right)=\iprod{\frac{\dd{f}}{\dd{\vec{\tau}_i}}}{\vec{\tau}_i},%
\end{align}
i.e.\ the surface divergence is the trace of tangential projection of the full gradient. For a differentiable tangential vector field $\vec{f}:\iface\mapsto T_\iface$, especially including the case $\vec{f}=\grads{f}$, the (surface) divergence theorem reads
\begin{align}
\int\limits_{\iface}{\divs{\vec{f}}\,\dd{o}}=\int\limits_{\partial\iface}{\iprod{\vec{f}}{\vec{n}_{\partial\iface}}\,\dd{l}}.\label{eqn:surface_divergence_theorem}%
\end{align}
For later application within the variational formulation, note that two scalar functions $f,g:\iface\mapsto\setR$ fulfill 
\begin{align}
\int\limits_{\iface}{g\laplaces{f}\,\dd{o}}=\int\limits_{\partial\iface}{g\iprod{\grads{f}}{\vec{n}_{\partial\iface}}\dd{l}}-\int\limits_{\iface}{\iprod{\grads{f}}{\grads{g}}\dd{o}},\label{eqn:trafo_variational_formulation}%
\end{align}
where the \name{Laplace-Beltrami} operator $\laplaces{}$ is introduced in subsection~\ref{subsec:laplace_beltrami_operator}. 
%
%
\section{Comparison of polynomials}\label{app:comparison_explicit_notation}%
The first three elements of the coefficient vector $\vec{\hat{u}}$, where $\hat{u}_{ij}$ corresponds to $t_1^{2i}t_2^{2j}\left(1+\curv_1^2t_1^2+\curv_2^2t_2^2\right)^{\frac{1}{2}}$, resulting from the polynomial comparison, cf.~subsection \ref{subsec:equating_polynomial_coefficients}, are
\begin{align}
%
%
\hat{u}_{00}=\frac{1}{3465D}&\left[%
72544885875\curv_1^{25} + 4193272314000\curv_1^{24}\curv_2 - 2577212786610\curv_1^{23}\curv_2^2 \right.\nonumber\\%
&- 26658698588694\curv_1^{22}\curv_2^3 - 57357128755944\curv_1^{21}\curv_2^4 + 336181054285530\curv_1^{20}\curv_2^5\nonumber\\%
&+ 41714627579527946\curv_1^{19}\curv_2^6 + 362007665932049430\curv_1^{18}\curv_2^7\nonumber\\%
&+ 1974693936115434347\curv_1^{17}\curv_2^8 + 8690216675233737542\curv_1^{16}\curv_2^9\nonumber\\%
&+ 18169798309454346220\curv_1^{15}\curv_2^{10} + 17997237348869644964\curv_1^{14}\curv_2^{11}\nonumber\\%
&+ 8400471266096489992\curv_1^{13}\curv_2^{12} + 7621880627263011268\curv_1^{12}\curv_2^{13}\nonumber\\%
&+ 18731291856940517620\curv_1^{11}\curv_2^{14} + 20016198248735780204\curv_1^{10}\curv_2^{15} \nonumber\\%
&+ 16084926436271498425\curv_1^9\curv_2^{16} + 4111980253565015324\curv_1^8\curv_2^{17}\nonumber\\%
&- 9815326789142415706\curv_1^7\curv_2^{18} - 6784297128026141358\curv_1^6\curv_2^{19}\nonumber\\%
&- 1393140036723497824\curv_1^5\curv_2^{20} - 146622851045506110\curv_1^4\curv_2^{21}\nonumber\\%
&- 12737928185209566\curv_1^3\curv_2^{22} + 250947833147550\curv_1^2\curv_2^{23}\nonumber\\%
&\left.+ 7442937203625\curv_1\curv_2^{24} + 5318812248750\curv_2^{25}\right],\\%
%
%
\hat{u}_{10}=\frac{-1}{2130D}&\left[%
72544885875\curv_1^{27} + 3830547884625\curv_1^{26}\curv_2 - 22099585675860\curv_1^{25}\curv_2^2\right.\nonumber\\%
&+ 64209322185606\curv_1^{24}\curv_2^3 - 206549792452944\curv_1^{23}\curv_2^4\nonumber\\%
&+ 1517363124557580\curv_1^{22}\curv_2^5 + 30461018702009246\curv_1^{21}\curv_2^6\nonumber\\%
&+ 109446918326844930\curv_1^{20}\curv_2^7 + 815480671742008547\curv_1^{19}\curv_2^8\nonumber\\%
&+ 1495391079214882517\curv_1^{18}\curv_2^9 + 4399306375420376470\curv_1^{17}\curv_2^{10}\nonumber\\%
&+ 11540162436209873564\curv_1^{16}\curv_2^{11} + 15876888881052595192\curv_1^{15}\curv_2^{12}\nonumber\\%
&+ 15758705774985074368\curv_1^{14}\curv_2^{13} + 9058115615694138220\curv_1^{13}\curv_2^{14}\nonumber\\%
&+ 14577128661181516004\curv_1^{12}\curv_2^{15} + 8744727214400259625\curv_1^{11}\curv_2^{16}\nonumber\\%
&+ 28905224435990940299\curv_1^{10}\curv_2^{17} + 2594680716262097144\curv_1^9\curv_2^{18}\nonumber\\%
&+ 7186618173791294142\curv_1^8\curv_2^{19} - 9947751822646978024\curv_1^7\curv_2^{20}\nonumber\\%
&- 6203435202180262860\curv_1^6\curv_2^{21} - 828643056358660266\curv_1^5\curv_2^{22}\nonumber\\%
&- 64844425789431750\curv_1^4\curv_2^{23} + 1826532320191425\curv_1^3\curv_2^{24}\nonumber\\%
&\left.- 63336193160625\curv_1^2\curv_2^{25} + 26594061243750\curv_1\curv_2^{26}\right],\\%
%
%
\hat{u}_{01}=\frac{-1}{2130D}&\left[%
362724429375\curv_1^{26}\curv_2 + 19594917775125\curv_1^{25}\curv_2^2 - 86674748460300\curv_1^{24}\curv_2^3\right.\nonumber\\%
&+ 146615450910390\curv_1^{23}\curv_2^4 - 1207840768860744\curv_1^{22}\curv_2^5\nonumber\\%
&+ 11196251748762756\curv_1^{21}\curv_2^6 + 252896928659490030\curv_1^{20}\curv_2^7\nonumber\\%
&+ 1200927891952953746\curv_1^{19}\curv_2^8 + 7556833261950904455\curv_1^{18}\curv_2^9\nonumber\\%
&+ 15745185870149404097\curv_1^{17}\curv_2^{10} + 15147291587893508942\curv_1^{16}\curv_2^{11}\nonumber\\%
&+ 10693380694498241020\curv_1^{15}\curv_2^{12} + 9860412201147581864\curv_1^{14}\curv_2^{13}\nonumber\\%
&+ 18073647507342869392\curv_1^{13}\curv_2^{14} + 13060950214817275468\curv_1^{12}\curv_2^{15}\nonumber\\%
&+ 26071491078811756420\curv_1^{11}\curv_2^{16} - 4777045933690144771\curv_1^{10}\curv_2^{17}\nonumber\\%
&+ 3674918930866985575\curv_1^9\curv_2^{18} - 9858935048252420176\curv_1^8\curv_2^{19}\nonumber\\%
&- 1260715003218935506\curv_1^7\curv_2^{20} - 727484776891384608\curv_1^6\curv_2^{21}\nonumber\\%
&- 577234908550047124\curv_1^5\curv_2^{22} - 81527477422926810\curv_1^4\curv_2^{23}\nonumber\\%
&- 14557017568197366\curv_1^3\curv_2^{24} + 319602838556925\curv_1^2\curv_2^{25}\nonumber\\%
&\left.- 19151124040125\curv_1\curv_2^{26} + 5318812248750\curv_2^{27}\right],%
\end{align}
with the common factor
\begin{align}
D=&\,942141375\curv_1^{28} + 50895890325\curv_1^{27}\curv_2 - 224230144725\curv_1^{26}\curv_2^2\nonumber\\%
&+ 598532006835\curv_1^{25}\curv_2^3 - 2428874075520\curv_1^{24}\curv_2^4 + 16349986503270\curv_1^{23}\curv_2^5\nonumber\\%
&+ 427125623057490\curv_1^{22}\curv_2^6 + 2231665401399570\curv_1^{21}\curv_2^7 + 14828318817700617\curv_1^{20}\curv_2^8\nonumber\\%
&+ 44864797857680423\curv_1^{19}\curv_2^9 + 119751294597828609\curv_1^{18}\curv_2^{10} + 214219532483111305\curv_1^{17}\curv_2^{11}\nonumber\\%
& + 237484900845042006\curv_1^{16}\curv_2^{12} + 227521098241347780\curv_1^{15}\curv_2^{13} + 193178681355432300\curv_1^{14}\curv_2^{14}\nonumber\\%
& + 273867280881223980\curv_1^{13}\curv_2^{15} + 191980286687434545\curv_1^{12}\curv_2^{16} + 373600162343077395\curv_1^{11}\curv_2^{17}\nonumber\\%
& + 26654117097834493\curv_1^{10}\curv_2^{18} + 94462186400242197\curv_1^9\curv_2^{19} - 112459641067919644\curv_1^8\curv_2^{20}\nonumber\\%
& - 58932688180872330\curv_1^7\curv_2^{21} - 30317166816236766\curv_1^6\curv_2^{22} - 10190296532147550\curv_1^5\curv_2^{23}\nonumber\\%
& - 1260751770087705\curv_1^4\curv_2^{24} - 183530830884975\curv_1^3\curv_2^{25} + 4003879094775\curv_1^2\curv_2^{26}\nonumber\\%
& - 164509592625\curv_1\curv_2^{27} + 69075483750\curv_2^{28}.%
\end{align}
\section{Parametrization of hypersurfaces using tesseral spherical harmonics}\label{app:parametrization_hypersurface_spherical_harmonics}%
Let $\mathbb{S}:=[0,2\pi)\times[0,\pi]$ be the parameter domain of the unit sphere in $\setR^3$. Then, the tesseral spherical harmonics $\mathcal{Y}_l^m:\mathbb{S}\mapsto\setR$ being defined as 
\begin{align}
\mathcal{Y}_l^m(\varphi,\theta)=\sqrt{\frac{2l+1}{4\pi}\frac{(l-|m|)!}{(l+|m|)!}}%
\begin{cases}
\sqrt{2}P_l^{|m|}(\cos\theta)\sin|m|\varphi&m<0\\%
\phantom{\sqrt{2}}P_l^m(\cos\theta)&m=0\\%
\sqrt{2}P_l^m(\cos\theta)\cos{m\varphi}&m>0\\%
\end{cases}
\end{align}
with the associated \name{Legendre} polynomials
\begin{align}
P_l^m(x)=\frac{(-1)^m}{2^ll!}\sqrt{(1-x^2)^m}\frac{\partial^{l+m}}{\partial x^{l+m}}(x^2-1)^l,%
\end{align}
form an orthonormal basis of the square-integrable functions $\mathcal{L}^2(\mathbb{S})$, where
\begin{align}
\delta_{lk}\delta_{mn}=\frac{1}{4\pi}\int\limits_{0}^{2\pi}{\int\limits_{0}^{\pi}{\mathcal{Y}_l^m\mathcal{Y}_k^n\sin\theta\dd{\theta}}\dd{\varphi}}. 
\end{align}
Within this paper, we consider a class of star-shaped hypersurfaces $\iface\subset\setR^3$ with parametrization 
\begin{align}
\iface=\{R\vec{e}_r:(\varphi,\theta)\in\mathbb{S}\},%
\end{align}
where $R:\mathbb{S}\mapsto\setR$ and $\vec{e}_r:=[\cos\varphi\sin\theta,\sin\varphi\sin\theta,\cos\theta]^{\sf{T}}$ denote the radius and radial unit vector, respectively. Herein, the third power of the radius instead of the radius itself is expressed in terms of spherical harmonics, i.e.
\begin{align}
R^3(\varphi,\theta)=\sum\limits_{l=0}^{L}{\sum\limits_{m=-l}^{l}{c_l^m\mathcal{Y}_l^m(\varphi,\theta)}}.\label{eqn:perturbed_sphere_radius_expansion}%
\end{align}
By recursive application of the contraction rule for spherical harmonics it can be shown that the order of $R$ is $\nicefrac{L}{3}$. However, the computation of the enclosed volume is considerably simplified, namely
\begin{align}
\lvert\operatorname{dom}(\iface)\rvert=\frac{1}{3}\int\limits_{0}^{2\pi}{\int\limits_{0}^{\pi}{\sum\limits_{l=0}^{L}{\sum\limits_{m=-l}^{l}{c_l^m\mathcal{Y}_l^m(\varphi,\theta)}}\sin\theta\dd{\theta}}\dd{\varphi}}=\frac{1}{3}\int\limits_{0}^{2\pi}{\int\limits_{0}^{\pi}{c_0^0\mathcal{Y}_0^0(\varphi,\theta)\sin\theta\dd{\theta}}\dd{\varphi}}=\frac{\sqrt{4\pi}}{3}c_0^0. 
\end{align}
Since this class of parametrizations degenerates at the poles, i.e. for $\theta\in\{0,\pi\}$, in order to ensure thet $\iface\in\mathcal{C}^0$ the derivative of the radius with respect to the azimuthal angle $\varphi$ needs to vanish, i.e.\ $\partial_\varphi R=0$ for $\theta\in\{0,\pi\}$. Then, the outer unit normal at the poles becomes
\begin{align}
\left.\ns\right\rvert_{\theta\in\{0,\pi\}}%
=\frac{R\vec{e}_r-\partial_\theta R\,\vec{e}_\theta}{\sqrt{R^2+(\partial_\theta R)^2}}.\label{eqn:spherical_parametrization_normal_pole}%
\end{align}
For \refeqn{spherical_parametrization_normal_pole} to be respectively unique obviously one requires the polar derivate to vanish at the poles as well, i.e.\ $\partial_\theta R=0$ for $\theta\in\{0,\pi\}$. While the tesseral spherical harmonics by definition fulfill $\partial_\varphi\mathcal{Y}_l^m\rvert_{\theta\in\{0,\pi\}}=0$, it holds that
\begin{align}
\left.\partial_\theta\mathcal{Y}_l^m\right\rvert_{\theta\in\{0,\pi\}}=%
\begin{cases}
0&|m|\neq1\\%
-\frac{1}{2}\sqrt{\frac{l(l+1)(2l+1)}{4\pi}}\cos^l\theta&|m|=1
\end{cases}
.%
\end{align}
Hence we exclude modes with $m=\pm1$ from the radius expansion, cf.~\refeqn{perturbed_sphere_radius_expansion}. For vanishing derivatives with respect to polar and azimuthal angle, the \name{Weingarten} map at the poles becomes
\begin{align}
\vec{W}=\frac{1}{R^2\sin^2\!\theta}\left[\begin{matrix}\sin\theta\,\partial_{\varphi\varphi}R-\sin^2\!\theta\,R&\partial_{\varphi\theta}R\\\sin^2\!\theta\,\partial_{\varphi\theta}R&-\sin^2\theta(R-\partial_{\theta\theta}R)\end{matrix}\right].%
\end{align}
Since by definition it holds that $\partial_{\varphi\varphi}R=\partial_{\varphi\theta}R=\partial_{\theta\theta}R=0$ for $\theta\in\{0,\pi\}$, the parametrization is sufficiently smooth at the poles with principal curvatures $\curv_i\rvert_{\theta\in\{0,\pi\}}=\nicefrac{-1}{R}$ and $\ns\rvert_{\theta\in\{0,\pi\}}=\vec{e}_r.$
\end{appendix}
\end{document}

%% file: lbquad_definitions.tex
\def\lvlset{\phi}%
\def\name#1{\textsc{#1}}%
\def\vec#1{\ensuremath{\bm{#1}}}%
\def\iprod#1#2{\ensuremath{\left\langle{#1,#2}\right\rangle}}%
\def\dyad#1#2{\ensuremath{#1\otimes#2}}%
\def\iface{\ensuremath{\Sigma}}%
\def\ifaceapprox{\ensuremath{\Gamma}}%
\def\grad#1{\ensuremath{\nabla{#1}}}%
\def\hessian#1{\ensuremath{\nabla^2{#1}}}%
\def\divs#1{\ensuremath{\operatorname{div}_\iface{#1}}}%
\def\trace#1{\ensuremath{\operatorname{tr}{#1}}}%
\def\laplace#1{\ensuremath{\Delta{#1}}}%
\def\curv{\ensuremath{\kappa}}%
\def\ident{\ensuremath{\vec{I}}}%
\def\PS{\ensuremath{\vec{P}_\iface}}%
\def\cell{\ensuremath{\mathcal{K}}}%
\def\face{\ensuremath{\mathcal{F}}}%
\def\triangle{\ensuremath{\mathcal{T}}}%
\def\ns{\ensuremath{\vec{n}_\iface}}%
\def\paradomain{\ensuremath{\mathcal{S}}}%
\def\dd#1{\ensuremath{\mathrm{d}#1}}%
\def\setR{\ensuremath{\mathbb{R}}}%
\def\setN{\ensuremath{\mathbb{N}}}%
\newcommand{\squareInt}[2][2]{\ensuremath{L^{#1}(#2)}}%
\newcommand{\fundet}[1]{\ensuremath{\mathcal{DF}(#1)}}%
\newcommand{\hilbert}[2][2]{\ensuremath{H^{#1}(#2)}}%
\newcommand{\reffig}[1]{figure~\ref{fig:#1}}%
\newcommand{\refFig}[1]{Figure~\ref{fig:#1}}%
\newcommand{\reftab}[1]{table~\ref{tab:#1}}%
\newcommand{\refeqn}[1]{eq.~(\ref{eqn:#1})}%
\newcommand{\testfun}[1][k]{\ensuremath{\varphi^{\mathrm{t}}_{#1}}}%
\newcommand{\testfunset}[1][N]{\ensuremath{\mathcal{F}^{\mathrm{t}}_{#1}}}%
\newcommand{\ansatzfun}[1][k]{\ensuremath{\varphi^{\mathrm{a}}_{#1}}}%
\newcommand{\ansatzfunset}[1][N]{\ensuremath{\mathcal{F}^{\mathrm{a}}_{#1}}}%
\newcommand{\error}{\ensuremath{\mathcal{E}}}%
\newcommand{\norm}[1]{\ensuremath{\left\lVert{#1}\right\rVert}}%
\newcommand{\grads}[2][\iface]{\ensuremath{\nabla_{#1}{#2}}}%
\newcommand{\laplaces}[2][\iface]{\ensuremath{\Delta_{#1}{#2}}}%

%% file: figures/tab/outofbounds.tex
\newcolumntype{C}[1]{>{\centering\arraybackslash}p{#1}} 
\begin{tabular}{c|*{3}{|C{0.75cm}|*{4}{C{0.5cm}}|}}%
\multicolumn{1}{c}{}&\multicolumn{5}{c}{$L=3$}&\multicolumn{5}{c}{$L=6$}&\multicolumn{5}{c}{$L=9$}\\
$N_\cell$&$N_\iface$&PG4&PG9&PC4&PC9&$N_\iface$&PG4&PG9&PC4&PC9&$N_\iface$&PG4&PG9&PC4&PC9\\\hline%
40&4820&0&0&0&0&4840&1&0&1&1&4996&1&0&1&0\\%
50&7524&0&0&0&0&7566&0&0&0&0&7772&0&1&0&1\\%
60&10870&1&1&1&0&10900&0&0&0&0&11250&1&2&2&1\\%
70&14790&0&0&0&0&14838&0&0&0&0&15324&0&0&1&2\\%
80&19302&0&0&0&0&19366&0&0&0&0&19992&1&2&1&3\\%
90&24450&0&0&0&0&24508&0&0&0&0&25290&0&0&0&0\\%
100&30163&0&0&0&0&30267&1&2&0&2&31206&0&1&0&1\\%
\end{tabular}%